\documentclass[useAMS,usenatbib]{mn2e}
\usepackage{times}
\usepackage{graphicx}
\usepackage{amsmath}
\usepackage{fmtcount}
\usepackage{amssymb}
\usepackage{gensymb}
\usepackage{natbib}
\usepackage{color}
\usepackage{booktabs}
\bibpunct{(}{)}{;}{a}{}{,}
\usepackage{psfig}
\title[The globular cluster system of NGC\,6861]{The extremely populated globular cluster system of the lenticular galaxy NGC\,6861}

\author[Escudero et al.]{Carlos G. Escudero$^{1,2,3}$, Favio R. Faifer$^{1,2,3}$, Lilia P. Bassino$^{1,2,3}$,\newauthor Juan Pablo Calder\'on$^{1,2,3}$, Juan Pablo Caso$^{1,2,3}$ \\
$^{1}$Facultad de Cs. Astron\'omicas y Geof\'isicas, UNLP, Paseo del Bosque S/N, 1900 La Plata, Argentina \\
$^{2}$ Instituto de Astrof\'isica de La Plata (CCT La Plata - CONICET - UNLP)\\
$^{3}$ Consejo Nacional de Investigaciones Cient\'ificas y T\'ecnicas, Rivadavia
1917, C1033AAJ Ciudad Aut\'onoma de Buenos Aires, Argentina
}

\begin{document}

\pagerange{\pageref{firstpage}--\pageref{lastpage}} \pubyear{}
\maketitle

\label{firstpage}

\begin{abstract}
We present a photometric study of the globular cluster (GC) system associated to 
the lenticular galaxy (S0) NGC\,6861, which is located in a relatively low density environment. 
It is based on GEMINI/GMOS images in the filters $g'$, $r'$, $i'$ of three fields, 
obtained under good seeing conditions. Analyzing the colour-magnitude and colour-colour 
diagrams, we find a large number of GC candidates, which extends out to 100 kpc, and we 
estimate a total population of $3000\pm300$ GCs.
Besides the well known blue and red subpopulations, the colour distribution 
shows signs of the possible existence of a third subpopulation with intermediate colours. 
This could be interpreted as evidence of a past interaction or fusion event. 
Other signs of interactions presented by the galaxy, are the non-concentric isophotes 
and the asymmetric spatial distribution of GC candidates with colours $(g'$--$i')_0$$>$1.16.
As observed in other galaxies, the red GCs show a steeper radial 
distribution than the blue GCs. In addition, the spatial distribution of these 
candidates exhibit strong signs of elongation. This feature is also 
detected in the intermediate subpopulation. On the other hand, the blue 
candidates show an excellent agreement with the X-ray surface 
brightness profile, outside 10 kpc. They also show a colour-luminosity 
relation (blue-tilt), similar to that observed in other galaxies.
A new distance modulus has been estimated through the blue subpopulation, which is in 
good agreement with the previous value obtained through the surface brightness fluctuations 
method.
The specific frequency of NGC\,6861 ($S_N=10.6\pm2.1$) is probably one of 
the highest values obtained for an S0 galaxy so far.
\end{abstract}


\section{INTRODUCTION}
\label{sec:intro}
It is known that globular clusters (GCs) are usually ancient stellar systems,  
with ages that can be established within a reliable and 
tight scale (e.g., \citealt{sala02,mend07}) that make them good tracers 
of the early evolutionary stages of their host galaxies.
From the observational point of view, GC systems reveal, to a 
greater or lesser extent, a bimodal colour distribution reaching, in some 
cases, a trimodal one (e.g. \citealt{blom12,caso13}). 
This multimodality indicates the presence of, at least, two subpopulations 
of GCs , usually referred as `blue' and `red' (e.g. \citealt{gebh99,lars01,kund01}).
They display different characteristics in terms of metallicity and projected 
spatial distribution, being the red GCs more metallic 
\citep{ushe12,schu10} and more concentrated towards their host galaxies 
than the blue ones.

These differences suggest that both subpopulations arose through two
different processes or in two phases. Among the most cited  scenarios 
about the origin and evolution of GC systems and their host galaxies, we 
find: (i) {\it the major merger 
formation scenario} \citep{ashm92} which proposes that the 
merger of two spiral galaxies (S) leads to an elliptical 
galaxy (E) with `blue' GC donated by the spirals and 
the `red' clusters formed in the merger; (ii) {\it the 
multiphase collapse scenario} \citep{forb97} which suggests that the 
`blue' GCs are formed first and their formation is 
truncated by the cosmic reionization \citep{cen01}, while the `red' 
GCs and the host galaxy field stars are formed later, in a second 
phase; (iii) {\it the accretion scenario} \citep{cote98}, in 
which `red' GCs are formed in a massive seed galaxy and 
the `blue' GCs are accreted from satellite galaxies of lower mass.
The inclusion of these scenarios in a proper cosmological framework 
(\citealt{beas02,pipi07,mura10,toni13}) has shown 
different degrees of success and failure in explaining the 
observed properties of GC systems. Therefore, it is reasonable to assume 
that the formation process of GC systems may include ingredients from all of them.

This close relationship between the galaxies and their associated GC systems, 
allows us to compare the predictions of different galaxy formation models 
with observational data. In this regard, we focus on the study of early-type 
galaxies and, in particular, of lenticular galaxies (S0) located in relatively low 
density environments. Most of the theories proposed for the origin of S0s are 
based on the results of different dynamical processes on normal spiral galaxies 
\citep{spit51}, such as interaction with the intracluster medium 
\citep{cowi77}, ram-pressure stripping (\citealt{bekk02,sun06}), 
galactic harassment \citep{moor96}, etc. The presence 
of S0 galaxies in low density environments can hardly be explained just by 
these mechanisms. Therefore, they represent a challenge to 
our understanding of galaxy formation and evolution.

In the last decade GCs were used to test the formation mechanism of S0 galaxies 
in a few works (\citealt{arag06,barr07}). However, up to date, the number of 
lenticular galaxies with GC systems studied in detail remains relatively low. 
Therefore, in order to draw more firm conclusions, it is necessary to 
enlarge the sample of S0 GC systems deeply analyzed. With that aims we 
started a study which is focused on several lenticular galaxies belonging 
to the field or low density groups.

In this context, we present here, as a first step, a study of the photometric 
properties of the GC system of the S0 galaxy NGC\,6861 (Table \ref{Tab0}). This 
galaxy is located in the Telescopium 
Group (AS0851) at a distance of 28.1 Mpc \citep{tonr01}, and is one 
of the brightest galaxies in the group together with the E galaxy 
NGC\,6868. Chandra observations show that NGC\,6861 might be
undergoing an interaction with NGC\,6868, suggesting the fusion of 
two sub-groups \citep{mach10}. 
NGC\,6861 presents an unusually high central stellar velocity 
dispersion of $\sim$414 km~s$^{-1}$ \citep{wegn03}, and from that \citet{mach10} 
inferred the presence of a supermassive black hole. Recently, \citet{rusl13} included 
this galaxy in their sample of ten massive early type galaxies. They found 
that NGC\,6861 (along with NGC\,4751, another S0 galaxy) shows the fastest 
rotation and the highest dispersion in their sample and confirmed the high 
dynamical mass for this black hole (see Table \ref{Tab0}).

The GC system of NGC\,6861 was previously studied by \citet{kund01}. They
 analyzed the GC systems of 29 S0 galaxies 
in the $V$ and $I$ bands (one image of 160 s in the F555W filter and two 
images of 160 s in the F814LP filter), observed between 1995 September 6 
and 1996 May 3 with the Wide Field and Planetary Camera 2 (WFPC2) of 
the Hubble Space Telescope. That work raises the possibility of the
existence of bimodality in the colour distribution of the GC 
candidates in the range 0.5 $<$$V$--$I$$<$ 1.5 mag.
Besides, the authors estimated a total GC population in their WFPC2 
field of 1858 GCs and a corresponding local specific frequency of 
$S_N=3.6\pm1.6$.

\section{DATA}
\subsection{Observations and Data Reduction}
\label{sec:obs}

\begin{figure}
\resizebox{0.99\hsize}{!}{\includegraphics{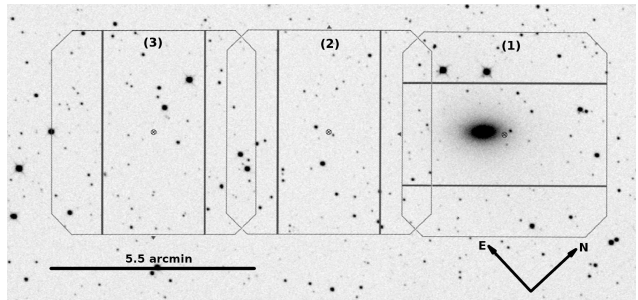}}
\caption{Digital Sky Survey image with the three superimposed GMOS fields.}
\label{fig1}
\end{figure}

We worked with images obtained in July-November 2010, using the Gemini 
Multi-Object Spectrograph \citep[GMOS,][]{hook04} of Gemini-South (Cerro 
Pachon, Chile) in imaging mode. The instrument consists of three 
2048 x 4608 pixel CCDs, separated by two 
$\sim$2.8 arcsec (39 pixel) gaps, with a scale of 0.0727 arcsec 
pixel$^{-1}$. The field of view (FOV) of the GMOS camera is 
$\sim 5.5 \times 5.5$ arcmin. Our images were taken under 
photometric conditions and using a $2\times2$ binning, which gives a 
scale of 0.146 arcsec pixel$^{-1}$ (Program GS-2010B-Q-2, PI: Bassino). 
Three deep fields were observed in the 
$g'$, $r'$ and $i'$ filters \citep{fuku96}, similar to 
those of the Sloan Digital Sky Survey (SDSS). The seeing conditions were 
excellent, ranging from 0.43 to 0.88 arcsec. As it 
can be seen in Figure\,\ref{fig1}, one of the fields was centred 
in NGC\,6861. In order to fill in the gaps and facilitate removal of cosmic 
rays and bad pixels, four slightly dithered exposures were taken per filter.

The raw data were processed by using GEMINI-GMOS routines within 
{\sc{iraf}}\footnote{IRAF is distributed by the National Optical Astronomical Observatories, 
which are operated by the Association of
Universities for Research in Astronomy, Inc., under cooperative agreement with the 
National Science Foundation} (e.g. {\sc{gprepare, gbias, giflat, gireduce}} and {\sc{gmosaic}}), and applying 
the appropriate bias and flat-field corrections. The bias and flat-field
images were acquired from the Gemini Science Archive (GSA) as part of the standard GMOS 
baseline calibrations.

It is known that $i'$ and $z'$ frames taken with GMOS show night-sky 
fringing. To subtract this pattern from our $i'$ data, it was necessary to create 
{\em fringe} calibration images by using the task {\sc{gifringe}}. These frames were built from 
seven $i'$ {\it blank sky} images taken with exposure times of 300 seconds. Subsequently, the 
night-sky fringing was removed from our $i'$ images using the task {\sc{girmfringe}}.

As a final step, the reduced images corresponding to the same filter were 
co-added using the task {\sc{imcoadd}}, obtaining $g'$, $r'$ and $i'$ co-added images for each 
field. These final images (see Table \ref{Tab1}) were then used for the subsequent photometric 
analysis.

\begin{table*}
\centering
\caption{Galaxy propierties. (2-5) Equatorial and galactic coordinates (NED); (6) morphological classification (NED); (7) B magnitude from the RC3 catalogue \citep{deva91}; (8) distance modulus from \citet{tonr01}; (9) effective radius from RC3; (10) central velocity dispersion calculated into a circular aperture of radius $r$$=$0.595 h$^{-1}$kpc \citep{wegn03}; (11) velocity dispersion within R$_e$ and (12) black hole mass from \citep{rusl13}.}
\label{Tab0}
\scriptsize
\begin{tabular}{cccccccccccc}
\toprule
\toprule
\multicolumn{1}{c}{\textbf{Galaxy}} &
\multicolumn{1}{c}{$\mathbf{\alpha}_{\mathbf{J2000}}$} &
\multicolumn{1}{c}{$\mathbf{\delta}_{\mathbf{J2000}}$} &
\multicolumn{1}{c}{\textbf{l}} &
\multicolumn{1}{c}{\textbf{b}} &
\multicolumn{1}{c}{\textbf{Type}} &
\multicolumn{1}{c}{\textbf{B${}_{\mathbf T}^{\mathbf 0}$}} &
\multicolumn{1}{c}{\textbf{(m--M)$_0$}} &
\multicolumn{1}{c}{\textbf{R$_e$}} &
\multicolumn{1}{c}{\textbf{$\sigma$$_0$}} &
\multicolumn{1}{c}{\textbf{$\sigma$$_e$}} &
\multicolumn{1}{c}{\textbf{M$_{BH}$}} \\
\multicolumn{1}{c}{} &
\multicolumn{1}{c}{\it {(h:m:s)}} &
\multicolumn{1}{c}{($^\circ:':''$)} &
\multicolumn{1}{c}{\it {($^\circ:':''$)}} &
\multicolumn{1}{c}{($^\circ:':''$)} &
\multicolumn{1}{c}{} &
\multicolumn{1}{c}{({\it mag})} &
\multicolumn{1}{c}{({\it mag})} &
\multicolumn{1}{c}{({\it $''$})} &
\multicolumn{1}{c}{({\it km s$^{-1}$})} &
\multicolumn{1}{c}{({\it km s$^{-1}$})} &
\multicolumn{1}{c}{({\it M$_{\odot}$})} \\
\midrule
NGC\,6861 & $20:07:19.5$ & $-48:22:13$ & $350:52:38$ & $-32:12:39$ & SA(s)0$^{-}$ & 11.92 & 32.24 & 17.7 & 414$\pm$17 & 388.8$\pm$2.6 & 2.2(2.1,2.4)x10$^9$ \\
\bottomrule
\end{tabular}
\end{table*}

\begin{table}
\centering
\caption{Observations of NGC\,6861, comparison field and blank sky. The values of airmass, exposure time and FWHM correspond to the final co-added images. Field 1 contains the galaxy.}
\label{Tab1}
\scriptsize
\begin{tabular}{lllll}
\toprule
\toprule
\multicolumn{1}{c}{\textbf{Galaxy}} &
\multicolumn{1}{c}{\textbf{Field}} &
\multicolumn{1}{c}{\textbf{Airmass}} &
\multicolumn{1}{c}{\textbf{T$_{exp.}$}} &
\multicolumn{1}{c}{\textbf{FWHM}} \\
\multicolumn{3}{c}{} &
\multicolumn{1}{c}{{(sec)}} &
\multicolumn{1}{c}{{(arcsec)}} \\
\multicolumn{2}{c}{} &
\multicolumn{1}{c}{{$g'$\hspace{4mm}$r'$\hspace{4mm}$i'$}} &
\multicolumn{1}{c}{{$g'$\hspace{5mm}$r'$\hspace{5mm}$i'$}} &
\multicolumn{1}{c}{{$g'$\hspace{3mm}$r'$\hspace{3mm}$i'$}} \\
\midrule
NGC\,6861    &~1  & 1.053~~1.056~~1.067  &  4x700~~~~4x350~~~~4x400  & 0.88~~0.75~~0.82  \\
             &~2  & 1.342~~1.251~~1.181  &  4x700~~~~4x350~~~~4x400  & 0.62~~0.48~~0.43  \\
	     &~3  & 1.272~~1.095~~1.134  &  4x700~~~~4x350~~~~4x400  & 0.57~~0.80~~0.77  \\
Comp. Field  &~-  & 1.026~~1.022~~1.039  & 14x180~~14x120~~17x90  & 0.56~~0.58~~0.53 \\
Blank sky    &~-  & ~~~~--{\hspace{4.5mm}}--{\hspace{2.8mm}}1.107  & ~~~~--{\hspace{6.5mm}}--{\hspace{4.1mm}}7x300  & ~~~--{\hspace{3.5mm}}--{\hspace{3.5mm}}-- \\
\bottomrule
\end{tabular}
\end{table}

\subsection{Photometry and classification of unresolved sources}
\label{sec:phot}

The source detection was performed on the $g'$ images because they present a 
better signal-to-noise ratio (S/N) than the $r'$ and $i'$ frames. In order to do that, we used 
a script that combines features of SExtractor \citep{bert96}, along with {\sc{iraf}} 
filtering tasks. It allows us to model and subtract both the background and the brightness 
due to the galaxy halo. 
To do this, initially, stellar objects were removed  by using SExtractor, and then the residual image was smoothed with a median filter, employing the {\sc{fmedian}} task from {\sc{iraf}}. This median filtered image was then subtracted from the original image. The procedure was applied a second time on the last image using a smaller median filter to detect and to discard weak halos around objects near the galactic centre \citep{faif11}.
For each field, the script provides a final catalogue with all 
detected objects and a galaxy light subtracted image.

From these lists of objects, $\sim$50 
bright unresolved sources were selected per field, uniformly distributed over the FOV. 
Then, second order models of the point spread function (PSF) were obtained using the {\sc{daophot}}
package \citep{stet87} within  {\sc{iraf}}. Several tests were performed with different psf models, 
selecting the moffat25 model, which gives lower fit errors than the Gaussian and moffat15 
options. The same procedure was applied to the $r'$ and $i'$ images of the same field. In 
order to obtain the instrumental psf magnitudes of all the detected sources, we applied the 
 {\sc{allstar}} task on each image, using their respective psf models and the object list of the 
reference filter ($g'$). In addition, we conducted an aperture correction to the psf magnitudes
through the  {\sc{mkapfile}} task.
As a final step, we built a master photometric catalogue in which the 
coordinates of objects belonging to different fields were in the same reference system. 
To do this, we selected common sources located in the overlapping regions of the fields and 
worked with the tasks  {\sc{geomap}} and  {\sc{geoxytran}} within  {\sc{iraf}}.

At the distance of NGC\,6861, we expect that GC candidates are unresolved sources. 
Therefore, the Stellarity Index of SExtractor (0 for resolved objects and 1 for 
unresolved ones) was used to make the object classification.  We set resolved/unresolved boundary in 0.5.

\subsection{Photometric Calibration}
\label{sec:ph_cal}

Several standard star fields observed during the same nights as our frames 
were downloaded from the GSA data base (see Table \ref{Tab2}). All of them 
are from the list of `Southern Standard Stars for the $u'g'r'i'z'$ System' of 
Smith et al. (in preparation)\footnote{ http://www-star.fnal.gov}. Exposure times 
for these images were 5.5 sec for the $g'$, $r'$ and 4.5 sec for the $i'$ filter. They 
were reduced using the same bias and flat-field images previously used 
for the reduction (see Section \ref{sec:obs}).

The {\sc{phot}} task within IRAF was used to obtain instrumental magnitudes of the 
standard stars through large apertures of  $\sim3$ arcsec of radius. Thus, no aperture 
corrections were applied to this photometry. Subsequently, and with the aim of 
bringing our instrumental magnitudes to the AB standard 
system, we used the transformation:

\begin{eqnarray}
\label{cero}
 m_{std} = m_{zero} + m_{inst} - K_{CP} (X-1)  + CT (m_1-m_2)_{std}
\end{eqnarray}

\noindent where $m_{std}$ are the standard magnitudes, m$_{zero}$ is the photometric 
zero point, $m_{inst}$ are instrumental magnitudes, K$_{CP}$ is the mean atmospheric 
extinction at Cerro Pach\'on given by the Gemini web 
page\footnote{http://www.gemini.edu/?q=node/10445}, $X$ is the airmass, CT is the 
colour term, and $(m_1-m_2)_{std}$ is one of the colours listed in the fourth 
column of Table \ref{Tab3}.

Following \citet{jorg09}, we applied the extinction correction and got a mean 
magnitude zero point of zero order for each standard frame, filter and night 
(values listed in Table \ref{Tab2}). After obtaining these zero points, we applied 
them to our science photometry and to the instrumental standard star 
magnitudes. Then, we used the full sample of these first order calibrated 
standard star magnitudes to get the residual zero point for every star:
\begin{eqnarray}
\Delta m_{zero}=m_{std}-[m_{inst}- K_{CP} (X-1) +m_{zero}]
\end{eqnarray}

\begin{figure*}
\resizebox{0.45\hsize}{!}{\includegraphics{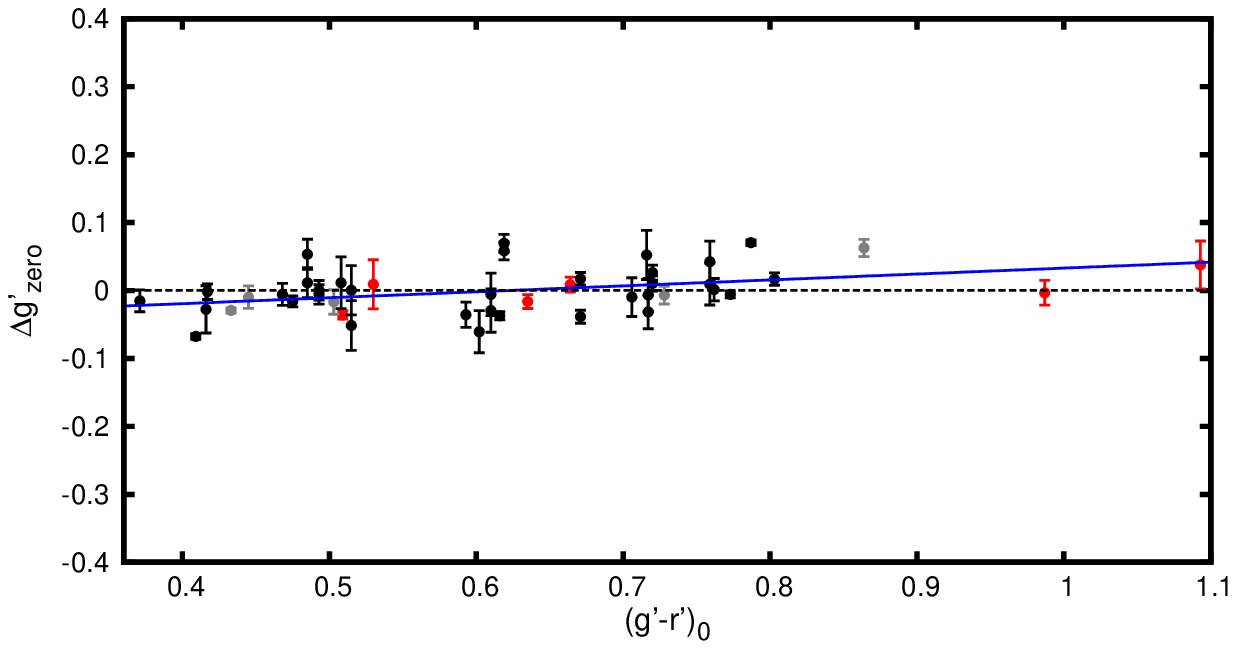}}
\resizebox{0.45\hsize}{!}{\includegraphics{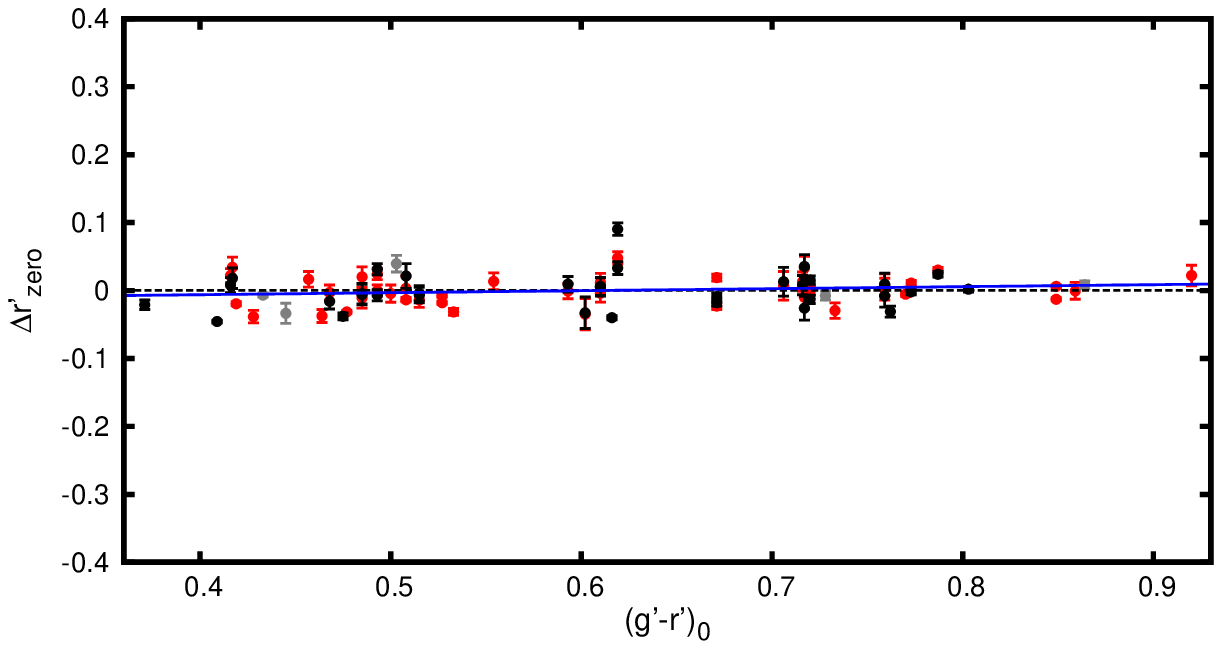}}
\resizebox{0.45\hsize}{!}{\includegraphics{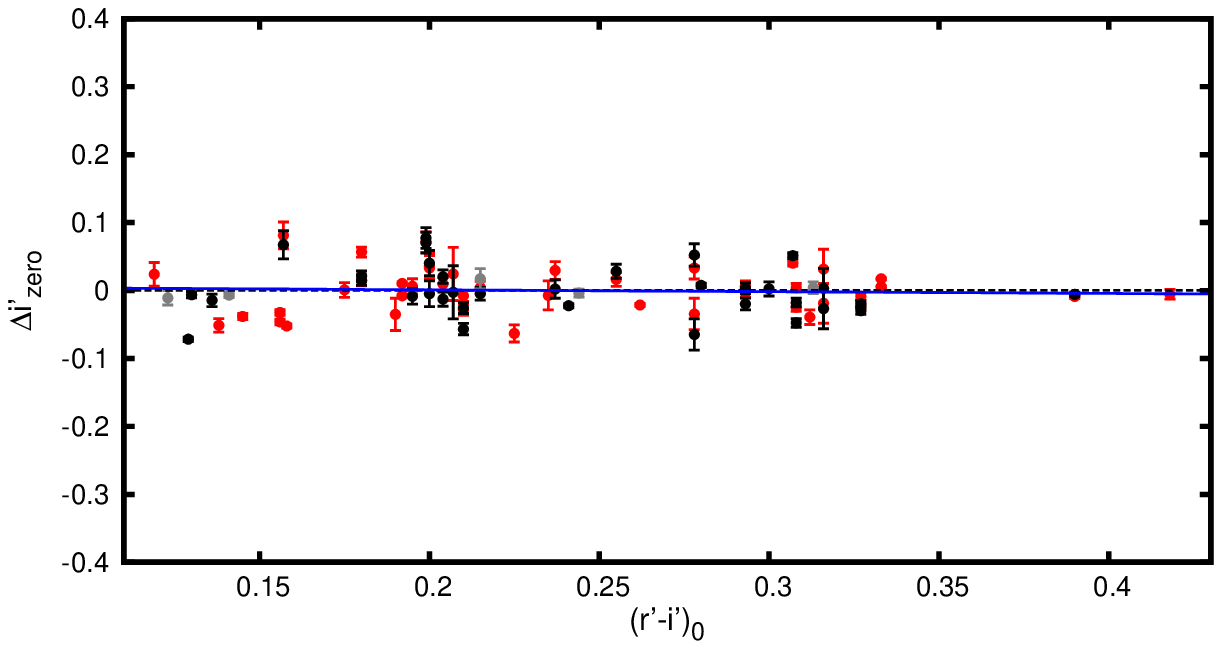}}
\caption{Residual zero points as a function of the standard colour index. Red, 
grey and black circles represent the photometric measurements performed on 
standard stars in the fields mentioned in Table \ref{Tab2}. The blue line 
shows the obtained fit.}
\label{fig3}
\end{figure*}

\noindent We fitted these `residual' zero points by using the following linear 
relation:  $\Delta m_{zero} = a + CT (m_1-m_2)_{std}$. Figure \ref{fig3} shows 
$\Delta m_{zero}$ versus the standard colours and the fitting functions. As can 
be seen in Table \ref{Tab3}, in the case of $g'$ and $r'$ we got 
significant values of $a$ and $CT$. However, the $i'$ parameters are 
consistent with zero.
Then, we applied the $g'$ and $r'$ colour terms to our previously 
calibrated science photometry.

Finally, objects in common in each science field were used to 
obtain any small zero point differences between our field 2 and the other 
pointings, resulting values lower than 0.03 mag in all cases. These small 
offsets 
were applied in order to refer our photometry to field 2. Subsequently, we visually 
rejected spurious detections and artifacts from our master catalogue 
(less than 1 per cent of total sources) and applied the Galactic 
extinction coefficients, given by \citet{schl11} ($A_g=0.179, A_r=0.124, A_i=0.092$).

\begin{table}
\caption{Standard fields used for the photometric calibrations. *S: number of standard stars in the field.
K$_{CP}$: atmospheric extinction coefficient at Cerro Pachon. 
m$_{zero}^{*}$: zero point obtained from the fit.}
\label{Tab2}
\scriptsize
\begin{tabular}{cccccc}
\toprule
\toprule
\multicolumn{1}{c}{\textbf{Standard field}} &
\multicolumn{1}{c}{\textbf{*S}} &
\multicolumn{1}{c}{\textbf{Science field}} &
\multicolumn{1}{c}{\textbf{Filter}} &
\multicolumn{1}{c}{\textbf{K$_{CP}$}} &
\multicolumn{1}{c}{\textbf{m$_{zero}^{*}$}} \\
\midrule
020000\_300600     &  2   &  1  &  $g'$  &  0.18  &  28.42$\pm$0.010  \\
220100\_300000 F1  &  4   &     &        &        &         \\
180000\_600000     &  15  &     &  $r'$  &  0.10  &  28.33$\pm$0.003  \\
180000\_600000 F5  &  16  &     &        &        &         \\
E8\_a F1           &  12  &     &        &        &         \\
180000\_600000     &  15  &     &  $i'$  &  0.08  &  27.92$\pm$0.004  \\
180000\_600000 F5  &  16  &     &        &        &         \\
E8\_a F1           &  12  &     &        &        &         \\
\midrule
180000\_600000     &  15  &  2  &  $g'$  &  0.18  &  28.41$\pm$0.006  \\
180000\_600000 F5  &  16  &     &        &        &         \\ 
220000\_595900 F1  &  4   &     &        &        &         \\
180000\_600000     &  15  &     &  $r'$  &  0.10  &  28.44$\pm$0.004  \\
180000\_600000 F5  &  16  &     &        &        &         \\ 
220000\_595900 F1  &  4   &     &        &        &         \\
180000\_600000     &  15  &     &  $i'$  &  0.08  &  28.01$\pm$0.006  \\
180000\_600000 F5  &  16  &     &        &        &         \\ 
220000\_595900 F1  &  4   &     &        &        &         \\
\midrule
020020\_600000     &  4   &  3  &  $g'$  &  0.18  &  28.24$\pm$0.016  \\
CDFS               &  2   &     &        &        &         \\
020020\_600000     &  4   &     &  $r'$  &  0.10  &  28.30$\pm$0.011  \\
CDFS               &  2   &     &        &        &         \\
020020\_600000     &  4   &     &  $i'$  &  0.08  &  27.92$\pm$0.005  \\
CDFS               &  2   &     &        &        &         \\
\bottomrule
\end{tabular}
\end{table}

\begin{table}
\centering
\caption{Residual zero point, colour terms and the final photometric 
zero points for field 2, obtained with the linear fits.}
\label{Tab3}
\scriptsize
\begin{tabular}{ccccc}
\toprule
\toprule
\multicolumn{1}{c}{\textbf{Filter}} &
\multicolumn{1}{c}{\textbf{a}} &
\multicolumn{1}{c}{\textbf{CT}}  &
\multicolumn{1}{c}{\textbf{(m$_1$--m$_2$)}} &
\multicolumn{1}{c}{\textbf{m$_{zero}$}} \\
\midrule
g' & -0.054$\pm$0.019 & 0.08$\pm$0.03 & $(g'$--$r')$ & 28.36$\pm$0.02  \\
r' & -0.018$\pm$0.011 & 0.03$\pm$0.01 & $(g'$--$r')$ & 28.42$\pm$0.01  \\
i' &  0.006$\pm$0.013 & -0.02$\pm$0.05 & $(r'$--$i')$ & 28.01$\pm$0.01 \\
\bottomrule
\end{tabular}
\end{table}

\subsection{Completeness test}
\label{sec:test}

To quantify a reliable detection limit of unresolved objects in our 
photometry, we performed a series of completeness tests in each filter and each field. We 
split the range 20.1 $<$$g'_0$$<$ 29.5 mag into intervals of 0.2 mag, adding 200 point sources 
using the {\sc{addstar}} task. A total number of 9600 artificial objects were added 
in each field. The spatial distribution of the artificial objects was 
carried out using a power law, trying 
to follow the radial distribution of the GC candidates. Subsequently, the same script 
mentioned in Section \ref{sec:phot} was applied to these images, in order to 
recover the added objects.

\begin{figure}
\resizebox{0.85\hsize}{!}{\includegraphics{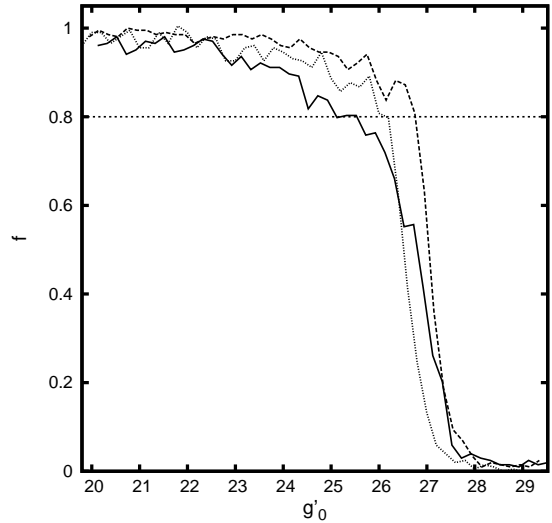}}
\caption{Completeness fraction as a function of $g'_0$ magnitude. The solid, dashed 
and dotted lines correspond to fields 1, 2 and 3, respectively. The horizontal dotted 
line shows the 80 per cent completeness level.}
\label{fig4}
\end{figure}

\begin{figure}
\resizebox{0.85\hsize}{!}{\includegraphics{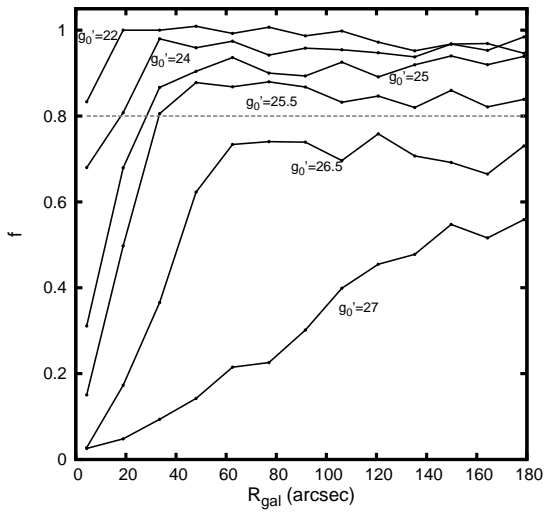}}
\caption{Completeness fraction as a function of galactocentric radius 
for field 1. The curves (from top to bottom) show the completeness for 
$g'_0$$=$$22, 24, 25, 25.5, 26.5, 27$ mag. The horizontal dashed line shows 
the 80 per cent completeness level.}
\label{fig5}
\end{figure}

Figure \ref{fig4} shows the completeness curves for the three fields, 
plotting the fraction of artificial objects recovered versus the input magnitude. It 
can be seen that, at $g_0'$$=$$25.5$ mag, the data have a completeness greater than 80 per cent. 
Thus, in the following analysis, we will take 
 a faint cut magnitude of $g_0'=25.5$ mag. Figure \ref{fig5} shows 
the completeness fraction for different ranges of magnitudes as a function of galactocentric 
radius ($R_{gal}$). It can be seen that most of the objects are lost at small $R_{gal}$, due 
to the high brightness of the galaxy. Taking this completeness variation into account, we can 
conclude that for magnitudes brighter than $g_0'$$=$$25.5$, we detect the vast majority 
of GCs over the whole NGC\,6861 field, except in the inner 30 arcsec 
of $R_{gal}$ (1.7 $R$$_e$) where most of the objects are lost.

\subsection{Comparison Field}
\label{sec:comp}

We were not able to get a comparison field as part of our Gemini program and, 
as we will mention in Section \ref{sec:dens}, our GC candidates seem to fill 
the whole field. Therefore, in order to estimate the background contamination, 
we had to search for deep $g'$, $r'$ and $i'$ fields in the GSA. We 
found only one nearby GMOS frame which filled our requirements. 
This field, centred in the compact galaxy group NGC\,6845, (RA=$20^h00^m58.28^s$, 
DEC=$-47^d04^m11.9^s$) is part of the Gemini program GS-2011A-Q-81 (PI: Gimeno). 
We downloaded the raw images and reduced and calibrated them according to 
the procedure explained in Subsections \ref{sec:obs} to \ref{sec:test}.

\begin{figure}
\resizebox{0.85\hsize}{!}{\includegraphics{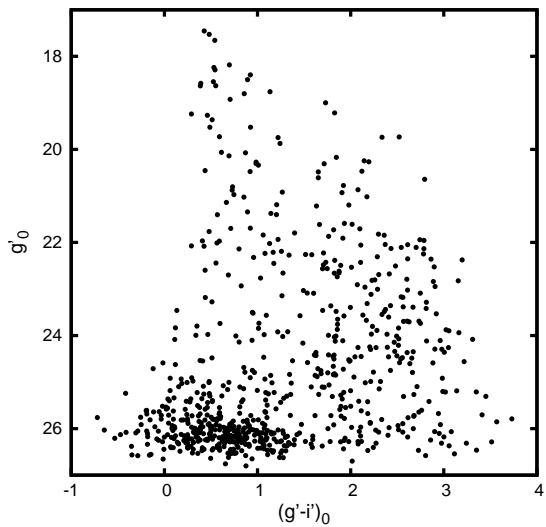}}
\caption{Colour-magnitude diagram of the points sources detected in the NGC 6845 field.}
\label{compAll}
\end{figure}

Unresolved objects were selected using the same ranges in colour and magnitude of our GC 
candidates in NGC\,6861 (see Section \ref{sec:color}). In Figure \ref{compAll} we show the 
colour-magnitude diagram for the unresolved sources obtained for this field. The completeness 
test on the $g'$ frame shows that we have a completeness $>$80 per cent for $g'$$<$ 26 mag. After that value, the completeness falls down strongly and colour errors grow rapidly.

\begin{figure}
\resizebox{0.85\hsize}{!}{\includegraphics{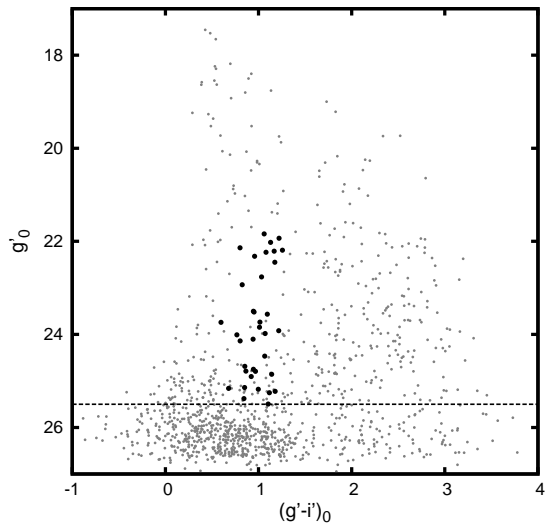}}
\caption{Colour-magnitude diagram of the points sources in the comparison 
field. Black filled circles show colours and magnitudes inside the ranges adopted for 
the GC candidates in NGC\,6861.}
\label{comp}
\end{figure}

It is important to mention that the NGC\,6845 group is composed of four 
galaxies. All of them are well contained inside the GMOS FOV. Two of these galaxies 
are spirals and two are of S0 type. The only available estimation of the distance 
to the group is based on the galaxy redshifts and they give values from 90 to 100 
Mpc \citep{gord03}. 
Adopting this range of distances and the $V$ to $g'$ band relation from 
\citet{faif11}, we can expect that massive GCs like $\omega~Cent$ will
show $g'_0\sim 24.8-25.1$ mag. Then, as a first step we 
should check if the GCs belonging to these galaxies are detected in 
our photometry. The colour-magnitude diagram shown in Figure\,\ref{compAll} 
indicates that no particular grouping is detected for objects in the colour 
ranges typical of GCs. We performed the same 
test using the colour-colour diagrams (CDs) and the results were 
similar. As an additional test we made plots of the spatial distribution 
of the point sources and only those objects with 
$g'_0$$>$24.5 mag seem to show some very marginal concentration 
towards the galaxies. Therefore, as expected from the assumed distance, some 
of the fainter objects in NGC\,6845 may well be GCs 
belonging to the galaxies in the group. 

We excluded the zones near to the four galaxies and we applied the same 
magnitude cuts as in the NGC\,6861 sample (see Section \ref{sec:color}) 
and thus only 36 point sources survive in the NGC\,6845 field 
(Figure\,\ref{comp}). This is a very low number and if we accept that 
some of them could be GCs belonging to the group, it means that 
our NGC\,6861 photometry has a contamination lower than 
8 per cent. In what follows we will take these 36 objects as our contamination
estimation.

\section{RESULTS}
\subsection{Galaxy surface profile and tidal structures}
\label{sec:elipse}

We obtained the surface brightness profile of NGC\,6861 using the 
{\sc{iraf}} task {\sc{ellipse}} which fits elliptical isophotes to the 
galaxy. In order to avoid the light contribution from nearby bright or 
extended objects, before executing {\sc{ellipse}} we masked them 
in the image. We modeled the galaxy light allowing the centre, ellipticity 
and position angle ($PA$) of the isophotes to vary freely until the fit 
became unstable in the outer regions. It is important to remember that the GMOS 
FOV is small and NGC\,6861 fills it completely. Therefore all the 
parameters mentioned here are limited to $R_{gal}$$<$ 2 arcmin.

\begin{figure*}
\resizebox{0.8\hsize}{!}{\includegraphics{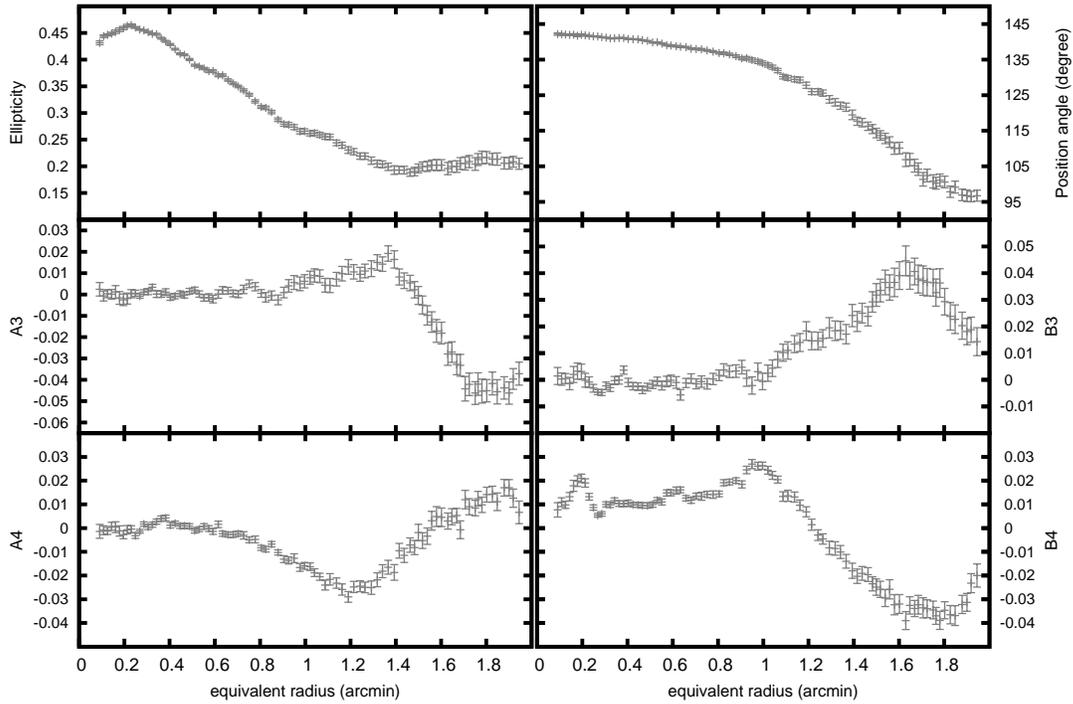}}
\caption{Isophotal parameters for the $g'$ filter versus equivalent 
galactocentric radius. The latter was calculated as the square root of 
the product of the semi-major and semi-minor 
axis lengths of the fitted ellipses. From left to right and from 
top to bottom, the graphics show the ellipticity and position angle of 
the isophotes, and the Fourier coefficients 
$A3, B3, A4, B4$, respectively.}
\label{fig2}
\end{figure*}

\begin{figure*}
\resizebox{0.4\hsize}{!}{\includegraphics{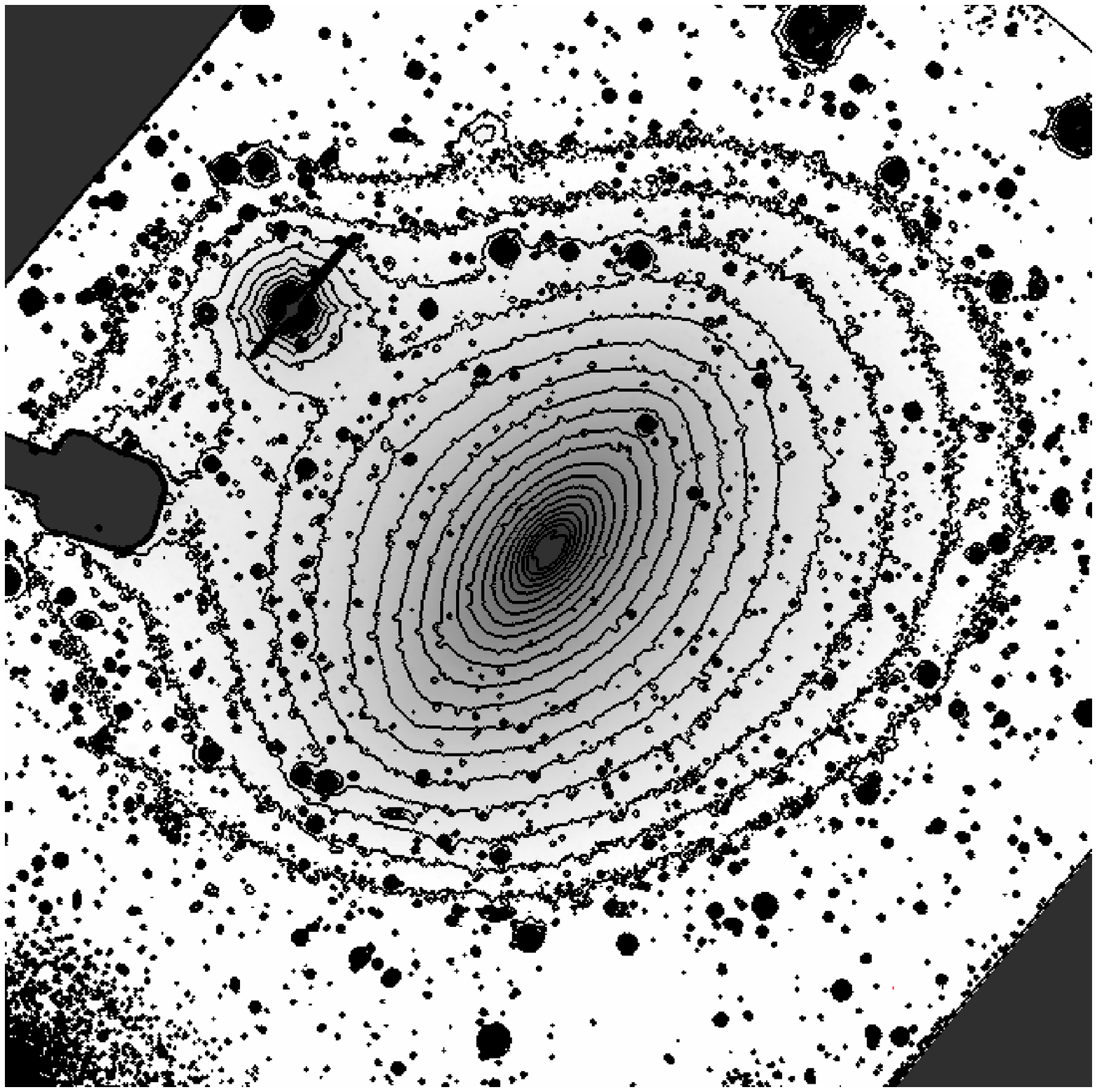}}
\resizebox{0.4\hsize}{!}{\includegraphics{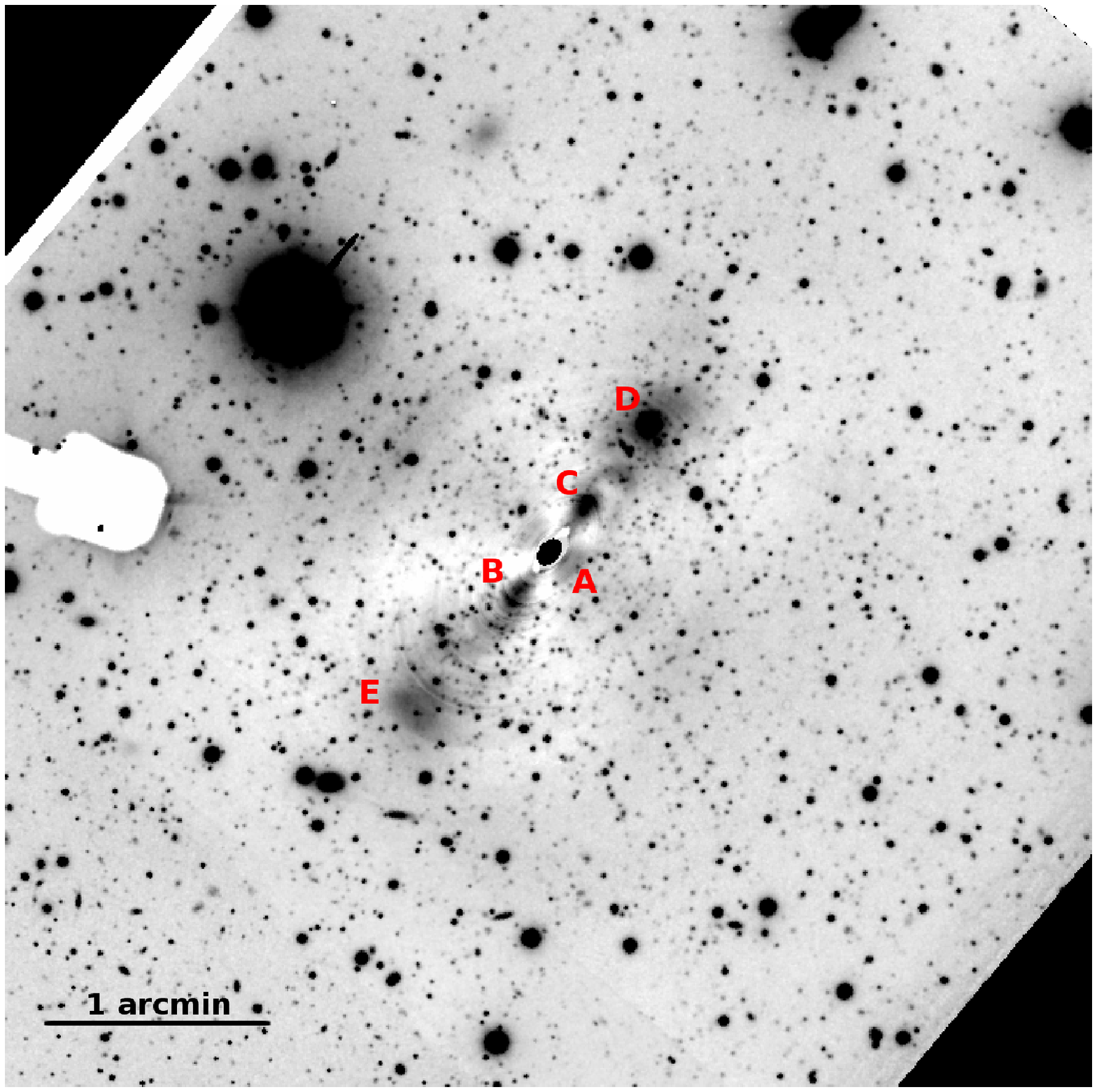}}
\caption{Left: $r'$ Isophotes of NGC\,6861 (field 1). The faintest and more 
external isophote correspond to $\mu_{r'}$$=$24.5 mag/arcsec$^2$, and brightest 
and inner to $\mu_{r'}$$=$17.5 mag/arcsec$^2$. The steps between them, are 
0.368 mag/arcsec$^2$. Right: Ratio between the the original $r'$ 
image and the model. It is possible to see the different structures present 
inside of NGC\,6861. 
The most obvious and easily visible structures were labelled by 
letters `A' to `E'. North is up and east is to the left.}
\label{nueva}
\end{figure*}

The results are shown in Figure\,\ref{fig2}. 
It can be seen that the ellipticity undergoes strong variations from 0.46 to 
0.2 within 100 arcsec, whereas the $PA$ shows a difference of $\sim$44\degree 
~between the innermost zone and the radius mentioned above. This ellipticity decrement 
is also seen in the X-ray hot gas emission from $e$$=$$0.4\pm0.04$ to $e$$=$$0.19\pm0.05$ for 
7.3$\leqslant$r$\leqslant$42 kpc \citep{mach10}.

These changes in the isophotal parameters are also detected in the 
Fourier terms ($A_3$,$B_3$,$A_4$,$B_4$), 
which measure the isophote's deviations from perfect ellipticity. 
The terms $A_3$ and $B_3$ give ``egg-shaped'' or ``heart-shaped'' 
isophotes, but the most significant term is the coefficient 
$B_4$ \citep{jedr87}. According to its sign, E/S0 
galaxies can be classified in disky ($B_4$$>$0) or boxy ($B_4$$<$0).
In our case, within 1 arcmin, $B_4$ indicates disky isophotes. At larger radii, the 
parameter $B4$ changes its sign showing that NGC\,6861 displays boxy isophotes. These 
results are in agreement with those obtained by \citet{li11} in the The Carnegie-Irvine 
Galaxy Survey (CGS). It is interesting to mention that boxy isophotes are usually 
suggestive of recent merging \citep{korm96}.

The variation shown by the $PA$ in NGC\,6861 was previously 
reported by \citet{tal09}. These authors mentioned that this galaxy has 
tidal characteristics, showing intra-group emission and non spherical isophotes. 
Therefore we searched this kind of evidence in our deep GMOS images. In the left 
hand panel of Figure \ref{nueva} we show the $r'$ image of NGC\,6861. 
It can be seen that the galaxy isophotes are not concentric, showing a 
deformation towards the N-W. The right hand panel in that figure shows the 
ratio between the original $r'$ image and the smooth ellipse model. 
Several distinct structures are evident in NGC\,6861, starting by a dust and 
stellar disk labelled with `A' ($R_{gal}$$<$10 arcsec), a series of more 
external arcs (where the brightest are labelled with `B' and `C'), and two 
diffuse stellar debris or low surface brightness structures. The feature 
located in the NW was labelled with `D' and that in the SE with `E'. Unfortunately, 
`D' presents a bright star on it. Between `C' and `D' there is clearly another 
structure. The structure `E', which is not symmetrically located with respect 
to `D', is easily visible and shows a sharp external edge. All of them 
lie along the inner major axis of the galaxy.

\begin{figure}
\resizebox{0.85\hsize}{!}{\includegraphics{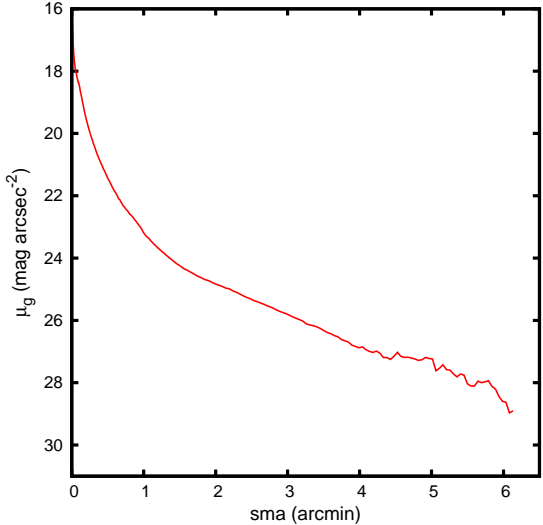}}
\caption{Radial surface brightness profile of NGC\,6861 in filter $g'$.}
\label{nueva2}
\end{figure}

Finally, and in order to compare the spatial light distribution of NGC\,6861 with that 
of the GCs, we obtained a more extended surface brightness profile for the galaxy. 
Although the images of fields 1 and 2 were taken on different days, 
Table \ref{Tab3} shows that the photometric conditions were similar. Therefore, we 
followed the same procedures as those included in the task {\sc{imcoadd}}, used 
the overlapping area of fields 1 and 2 to equalize the signal and the sky 
level of the $g'$ images, and built a mosaic including these two fields. After 
checking the photometric scales by comparing common objects in the original 
images with those in the final mosaic, we run {\sc{ellipse}} and we get a 
profile reaching 6 arcmin of $sma$ (Figure \ref{nueva2}). The inner zone of 
the profile ($sma$$<$70 arcsec), which is saturated in our long exposure 
images, was obtained by running {\sc{ellipse}} on a short exposure $g'$ image.

\subsection{Globular cluster colours}
\label{sec:color}

\begin{figure*}
\resizebox{0.33\hsize}{!}{\includegraphics{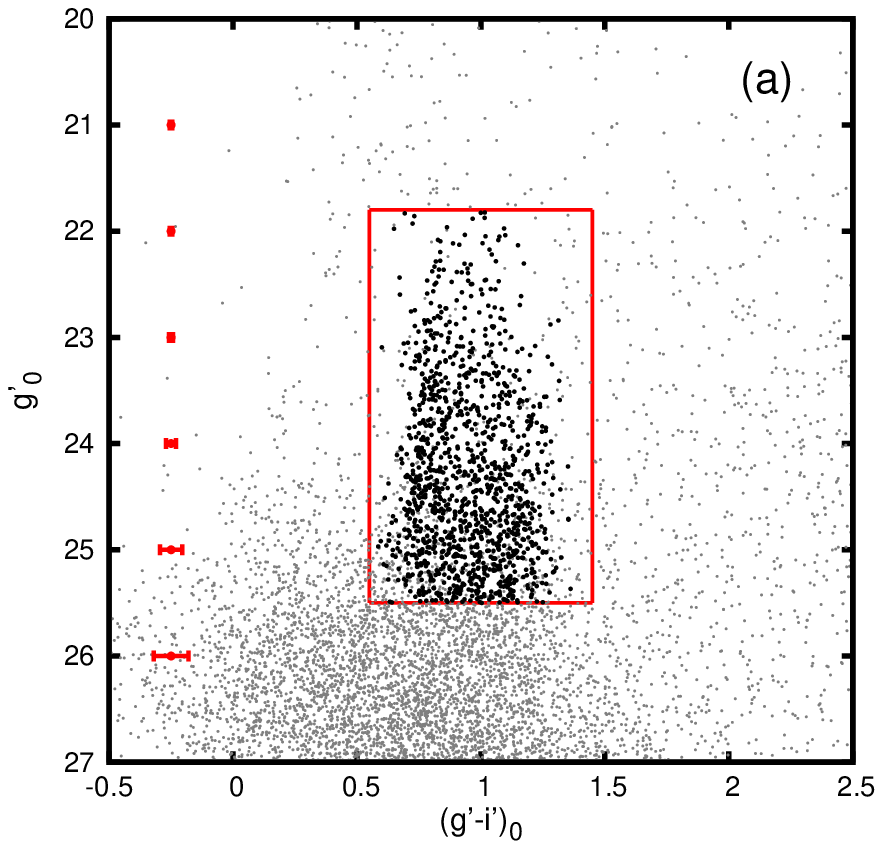}}
\resizebox{0.33\hsize}{!}{\includegraphics{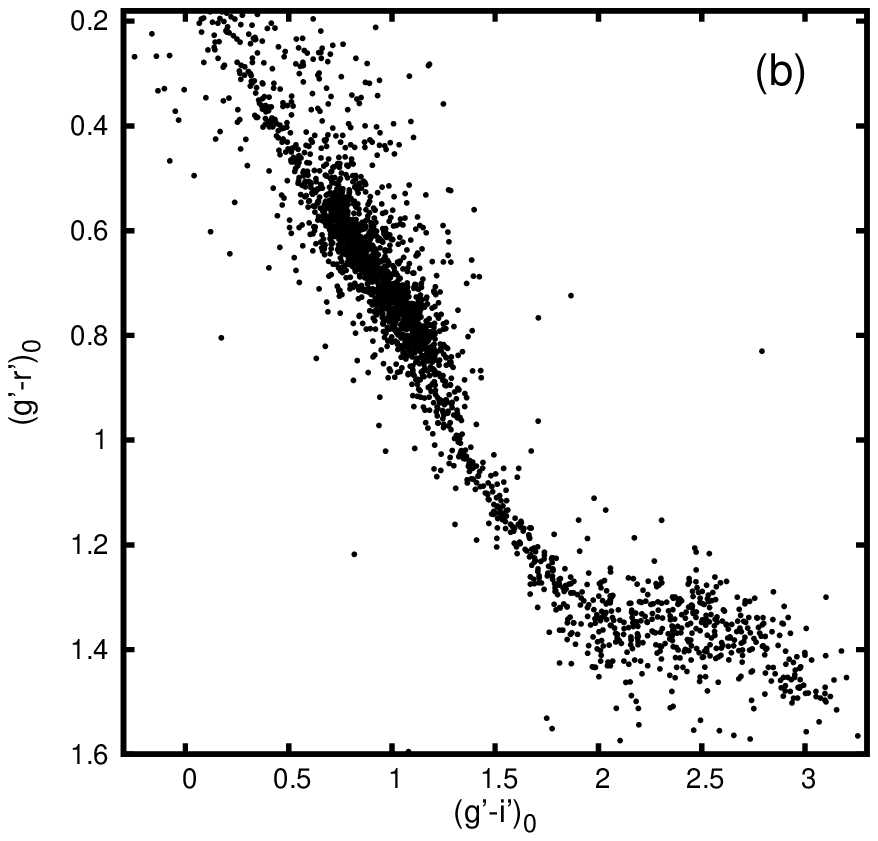}}
\resizebox{0.33\hsize}{!}{\includegraphics{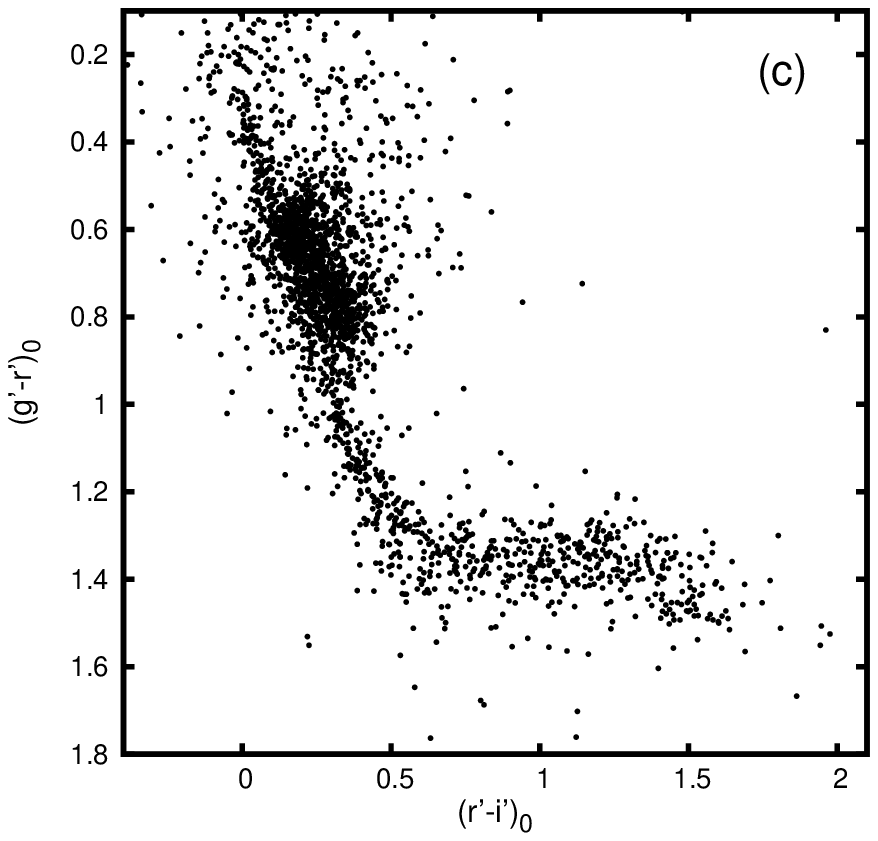}}
\caption{Colour-magnitude (a) and colour-colour diagrams (b,c) of all of
the detected point sources in the three fields (black dots), and the 
initial sample of GC candidates (black filled circles). Mean photometric 
errors $(g'$--$i')_0$ are shown in panel (a) by red bars. The red box shows 
the limits used in colour and magnitude for the selection of GC candidates.}
\label{fig6}
\end{figure*}

Figure\,\ref{fig6} shows  the $g'_0$ versus $(g'$--$i')_0$ colour-magnitude 
diagram (CMD), as well as the $(g'$--$r')_0$ versus $(g'$--$i')_0$ and 
the $(g'$--$r')_0$ versus $(r'$--$r')_0$ colour-colour diagrams, 
which include all the unresolved sources detected in the three GMOS fields. 
As it can be seen, GCs are easily identified as point sources clustered 
around specific colours and most of these unresolved objects define a clear 
sequence, which is highlighted 
above the sequence of the Milky Way stars. In order to obtain our sample 
of GC candidates, we selected the objects within the colour ranges 
0.5 $<$$(g'$--$r')_0$$<$ 0.95 mag; 
0.05 $<$$(r'$--$i')_0$$<$ 0.5 mag; 0.55 $<$$(g'$--$i')_0$$<$ 1.45 mag. 
We expect that a small percentage of the GC candidates are foreground 
stars and possible Ultra Compact Dwarfs \citep[UCDs;][]{drin03}.
To separate bright GCs from UCDs and/or field stars, we took $g_0'$$=$21.8 
mag as the bright end in the CMD, which is roughly  
equivalent to $M_V\sim-11$ mag. 
(in agreement with the value suggested by \citealt{mies06}). The 
adopted faint end was $g_0'$$=$25.5 mag, where we ensure a completeness 
level greater than 80 per cent, and colour photometric errors lower than 
0.1 mag.

\begin{figure}
\resizebox{0.95\hsize}{!}{\includegraphics{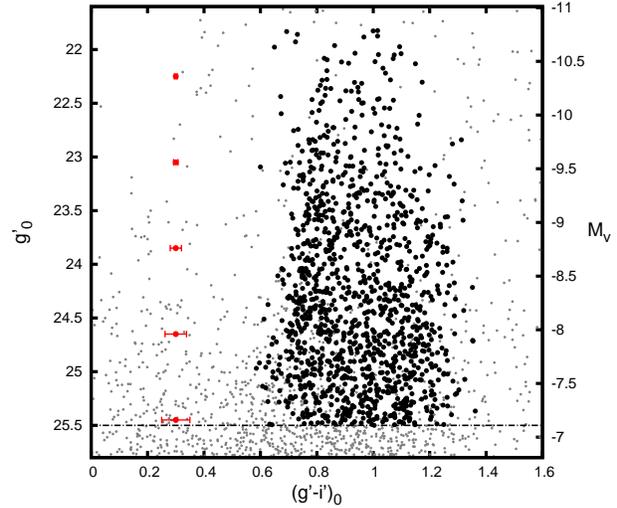}}
\caption{Colour-magnitude diagram of the final GC candidates (black filled circles), the dashed line 
shows the 80 per cent completeness level.}
\label{fig7}
\end{figure}

The final sample of GC candidates for the subsequent analysis is 
shown in Figure\,\ref{fig7}. A total of 1245 objects meet 
the above criteria. The figure shows all the candidates identified 
before performing any statistical subtraction of field contamination. 

At first glance, the CMD seems to show bimodality, displaying a marked blue 
subpopulation and a more disperse, but large, red subpopulation. Interestingly, 
Figure\,\ref{fig7} shows that the blue sequence of GCs seems to become redder with 
increasing luminosity. This particular characteristic shown by NGC\,6861, is
known as `blue-tilt'. We will discuss it in Subsection \ref{sec:tilt}.

\subsection{Colour histograms}
\label{sec:histo}

\begin{figure}
\resizebox{0.85\hsize}{!}{\includegraphics{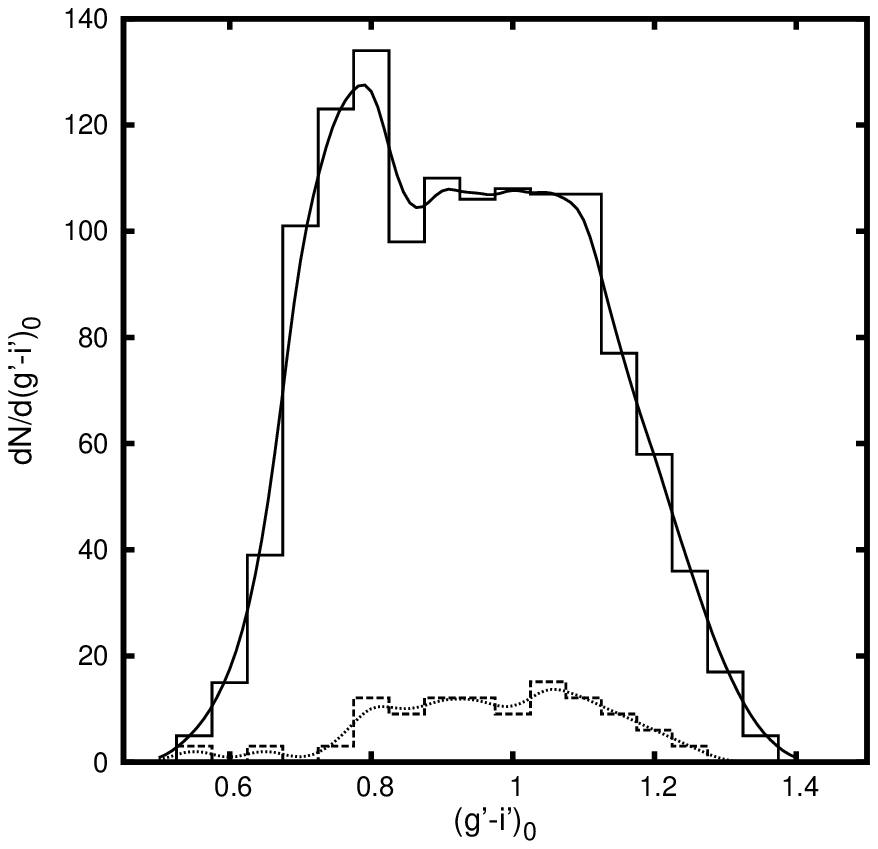}}
\resizebox{0.85\hsize}{!}{\includegraphics{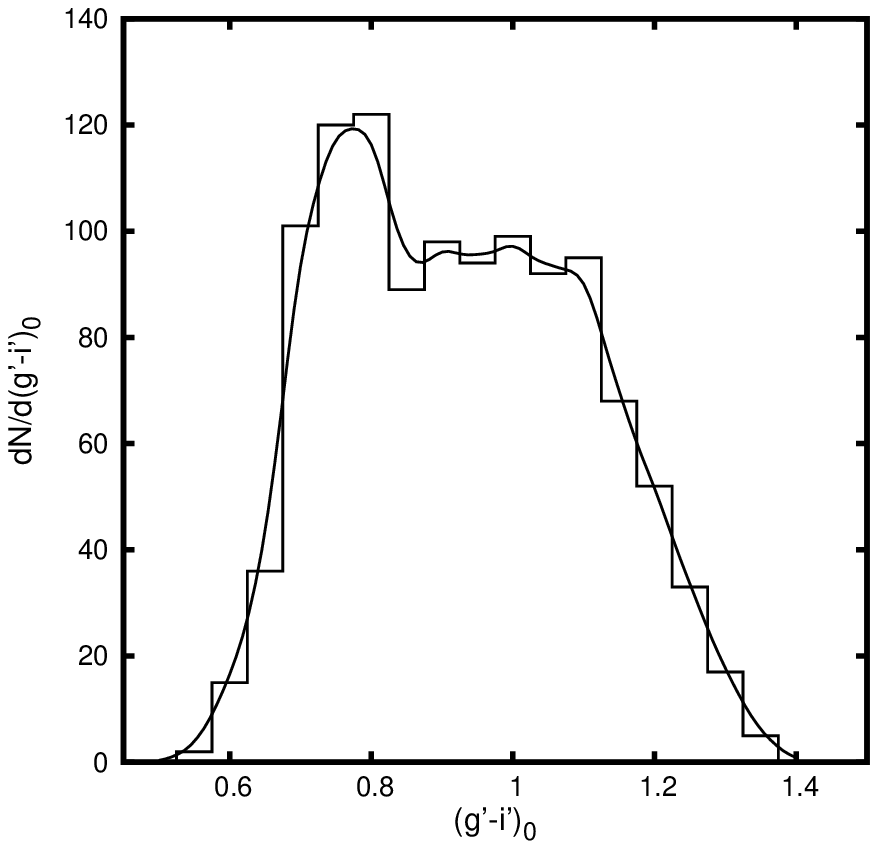}}
\caption{{\it Upper panel:} $(g'$--$i')_0$ colour histograms for the GC candidates and comparison field (dashed lines) with magnitudes between 21.8 $<$$g_0'$$<$ 25.5. The solid and dotted lines represent the smoothed colour distributions. {\it Bottom panel:} Background corrected colour histogram and smoothed distribution.}
\label{fig8}
\end{figure}

Due to the large number of GC candidates, we were able to
build colour histograms counting objects in small bins of 0.05 mag. 
In the upper panel of Figure\,\ref{fig8}, we show the histogram 
obtained for all candidates in the sample and the expected 
contamination obtained, as it was indicated in 
Subsection\,\ref{sec:comp}. A factor of 3.0 was applied to the 
contaminant object counts in order to take into account the 
different areas of the NGC\,6861 and NGC\,6845 fields. As it was mentioned in 
that section, the number of MW stars and background galaxies in the GCs
sample seems to be very low. Therefore, in the 
lower panel of Figure\,\ref{fig8} we show the histogram corrected 
by the background contamination and, as expected, the 
correction had a negligible effect on the integrated colour 
distribution. In the same figures we show the respective smoothed colour 
distributions obtained using a Gaussian kernel of $\sigma$$=$0.04 mag. 
This $\sigma$ value was considered to be representative of the 
mean error in $(g'$--$i')_0$ colours for fainter candidates.

To quantify the modes and dispersions of the different subpopulations of GC 
candidates, we used the statistical algorithm Gaussian Mixture 
Model \citep[GMM;][]{mura10}. 
This method is a parametric probability density function 
represented as a weighted sum of Gaussian component 
densities. Using three statistics, which are 1) parametric
 bootstrap method (low values indicate a multi-modal distribution), 2) 
separation of the peaks between Gaussians (D$>$2 implies multi-modal 
distribution) and 3) kurtosis of the 
input distribution (k$<$0 condition necessary but not sufficient for 
bimodality), GMM quantifies whether a multi-modal distribution 
provides a better fit 
than a unimodal one. It also allows to define initial values 
for the modes and fits equal dispersions (homoscedastic) or different 
dispersions (heteroscedastic) to the components. 

We started by fitting two Gaussian distributions and finding mean colours 
for the expected `classic' blue and red GCs. The brightest 
objects ($g'_0$$<$22.8 mag) were rejected in order to avoid the effect 
of the `blue tilt' and the usual unimodal colour distribution of 
these massive objects. The values obtained, listed in 
Table\,\ref{Tab4}, are in good agreement with the location of 
the peaks found in other galaxies by \citet{faif11}. In that work, the 
authors present results for another bulge dominated S0 galaxy, 
NGC\,3115, which is considered
now as one of the most clear examples of bimodality in 
colour ($\langle$$g'$--$i'$$\rangle_0$$\sim$0.765$\pm$0.007 
and $\langle$$g'$--$i'$$\rangle_0$$\sim$1.044$\pm$0.011 for blue 
and red GCs, respectively) and in metallicity \citep{brod12}. 
Interestingly, NGC\,3115 shows more or less the same luminosity 
as NGC\,6861, but the latter presents a larger GC 
population and its bimodality is not as clear.

\begin{figure}
\resizebox{0.51\hsize}{!}{\includegraphics{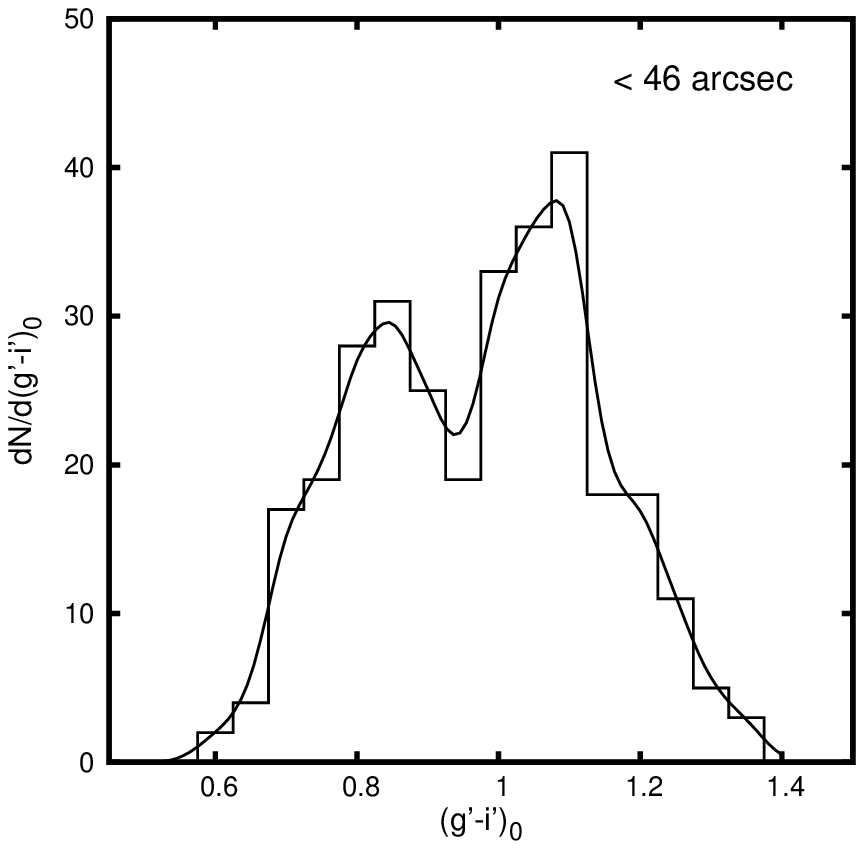}}
\resizebox{0.51\hsize}{!}{\includegraphics{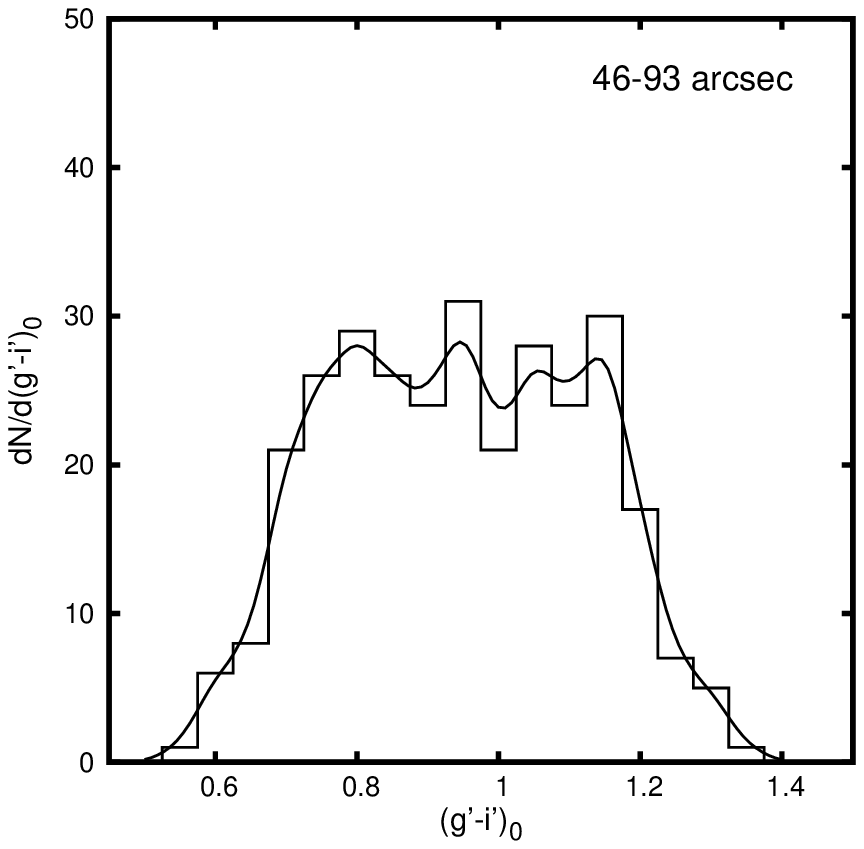}}
\resizebox{0.51\hsize}{!}{\includegraphics{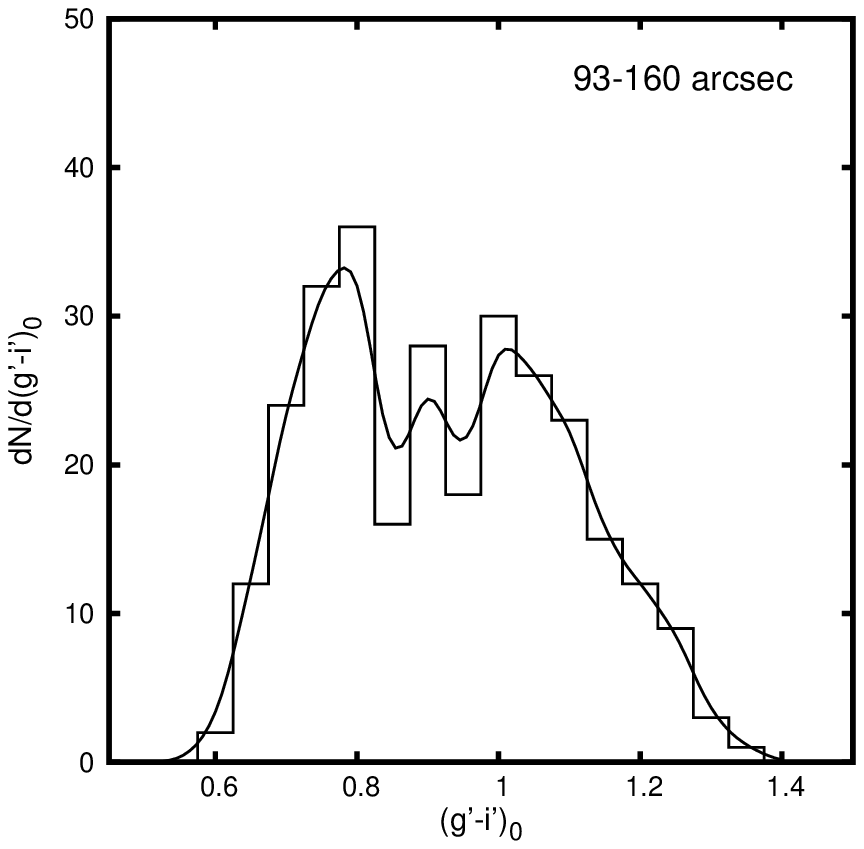}}
\resizebox{0.51\hsize}{!}{\includegraphics{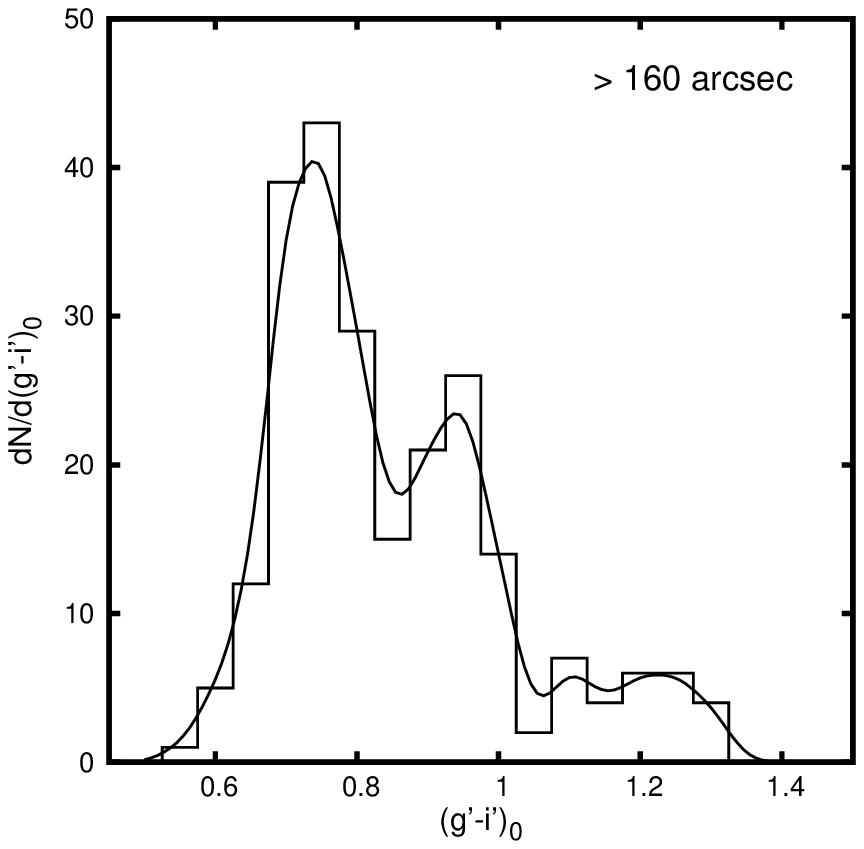}}
\caption{Colour distribution of the GC candidates, separated in four radial bins, according 
to their galactocentric distance (0-46, 46-93, 93-160 and $>$160 arcsec). The 
solid line represents the applied Gaussian kernel.}
\label{fig9}
\end{figure}

Figure\,\ref{fig8} shows that the red subpopulation 
in NGC\,6861 presents a broad and flat colour distribution. 
\citet{blom12} found similar characteristics in the massive 
elliptical galaxy NGC\,4365, where the existence of a third 
subpopulation of GCs with intermediate colours is proposed. 
In order to study whether the same situation occurs in NGC\,6861, we split the 
sample of GC candidates in four radial bins with approximately 310 objects 
in each of them ($R_{gal}\leq$ 46, 46-93, 93-160 and $>$160 arcsec). 
The background-corrected colour distributions are shown in Figure\,\ref{fig9}. 
A Gaussian kernel was used in order to obtain smoothed colour 
distributions, as described earlier.
We can see that in the inner region there is a clear blue 
subpopulation at modal value $(g'$--$i')_0\sim0.85$. Besides this, 
a clearer red peak is easily detected at modal colour of $(g'$--$i')_0\sim1.10$. As we move 
towards outer regions, and as expected, the blue GCs begin to dominate. 
However, the red GC candidates become more 
heterogeneous and at least two other possible peaks seem 
to appear. One of them in $(g'$--$i')_0\sim0.94$ and the other one 
around $(g'$--$i')_0\sim1.10-1.15$ mag.

With the purpose of quantifying and characterizing the possible presence of 
three subpopulations, we ran GMM to the colour distribution and the 
RMIX\footnote{RMIX is publicly available at 
http://www.math.mcmaster.ca/peter/mix/mix.html} software to the 
histogram (whole sample background-corrected histogram), 
fitting in both cases a trimodal distribution. The obtained values for the 
GMM fit, listed in Table\,\ref{Tab4}, indicate that a trimodal distribution is
as good as a bimodal one. On the other hand, the reduced chi-square
 values obtained with RMIX indicate that a trimodal case gives
a better fit to the data than the bimodal one.

\begin{figure}
\resizebox{1.0\hsize}{!}{\includegraphics{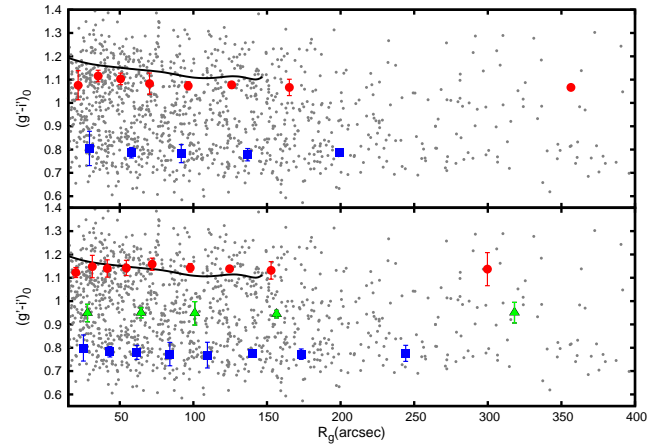}}
\caption{Colour index $(g'$--$i')_0$ vs. projected galactocentric radius. The 
bimodal case is plotted in the upper panel and the trimodal case in the 
bottom panel. Red circles, blue squares and green 
triangles represents mean values in colour for the red, blue and 
intermediate subpopulations, respectively. 
The black line show the profile colour of the galaxy 
within a 2.5 arcmin radius.}
\label{fig10}
\end{figure}

In different early-type galaxies, the mean colour of the red GCs
were found to be similar to the inner halo colour of their host 
galaxies \citep{forb01}. This is the consequence of the 
bulge/spheroid field stars and red GCs having a strong genetic 
nexus \citep{fort09}. Through the task {\sc{ellipse}} we obtained 
an approximate colour profile for NGC\,6861 (see Section 
\ref{sec:elipse}). This profile was calculated within a radius 
less than 2 arcmin, in order to avoid the strong effect of the error in 
the sky level. In the upper panel of Figure\,\ref{fig10} we compare 
the colour of the galaxy halo with the mean colour of the GCs for 
the bimodal case.
To separate blue and red subpopulations, we consider the colour cut 
in $(g'$--$i')_0= 0.90$.
It can be seen that the red GCs look similar to the inner halo, although 
somewhat bluer. However, considering a trimodal case 
($(g'$--$i')_0\leq$0.88, 0.88$<$$(g'$--$i')_0$$<$1.01, and 
$(g'$--$i')_0\geqslant$1.01, for 
blue, green and red subpopulations, respectively), we can see that the red 
GCs look almost identical in colour to the halo (lower panel in 
Figure \ref{fig10}). In both cases, the GC subpopulations were split using 
the values provided by GMM, and subsequently mean colours were obtained 
within several galactocentric intervals containing the same number 
of objects, 85 GCs in the bimodal case and 51 in the trimodal one.
It is interesting to see that figure 16 in \citet{blom12} 
presents similar results.

\begin{table*}
\centering
\caption{Values obtained with GMM and RMIX for the $(g'$--$i')_0$ colour 
distribution. Statistics values ($p$, $D$, $k$, $\chi^2$) are listed 
along with the mean colour $\mu$, their dispersion $\sigma$ and the 
fraction of objects assigned by RMIX for each subpopulation for the 
bimodal and trimodal case.
}
\label{Tab4}
\scriptsize
\begin{tabular}{lcccccccccc}
\toprule
\toprule
\multicolumn{1}{c}{\textbf{}} &
\multicolumn{4}{c}{\textbf{GMM}} &
\multicolumn{6}{c}{\textbf{RMIX}} \\
\multicolumn{1}{c}{\textbf{Population}} &
\multicolumn{2}{c}{\textbf{Bimodal case}} &
\multicolumn{2}{c}{\textbf{Trimodal case}} &
\multicolumn{3}{c}{\textbf{Bimodal case}} &
\multicolumn{3}{c}{\textbf{Trimodal case}} \\
\multicolumn{1}{c}{\textbf{Statistics values}} &
\multicolumn{2}{c}{\text{($p$=0.01; $D$=2.63$\pm$0.14; $k$=-0.934)}} &
\multicolumn{2}{c}{\text{($p$=0.01; $D$=2.46$\pm$0.34; $k$=-0.934)}} &
\multicolumn{3}{c}{\text{($\chi^2$=15.639)}} &
\multicolumn{3}{c}{\text{($\chi^2$=5.869)}} \\
\multicolumn{1}{c}{} &
\multicolumn{1}{c}{\textbf{$\mu$}} &
\multicolumn{1}{c}{\textbf{$\sigma$}} &
\multicolumn{1}{c}{\textbf{$\mu$}} &
\multicolumn{1}{c}{\textbf{$\sigma$}} &
\multicolumn{1}{c}{\textbf{$\mu$}} &
\multicolumn{1}{c}{\textbf{$\sigma$}} &
\multicolumn{1}{c}{\textbf{$f$}} &
\multicolumn{1}{c}{\textbf{$\mu$}} &
\multicolumn{1}{c}{\textbf{$\sigma$}} &
\multicolumn{1}{c}{\textbf{$f$}} \\
\midrule
Blue   & 0.79$\pm$0.02 & 0.08$\pm$0.007 & 0.783$\pm$0.014 & 0.075$\pm$0.009 & 0.783$\pm$0.01  & 0.07$\pm$0.006  & 0.376$\pm$0.023 & 0.778$\pm$0.02 & 0.07$\pm$0.003 & 0.379$\pm$0.025 \\
Green  &    -           &       -        & 0.954$\pm$0.062 & 0.066$\pm$0.032 &      -          &     -           & - & 0.955$\pm$0.03 & 0.07$\pm$0.020 & 0.227$\pm$0.140 \\
Red    & 1.07$\pm$0.02 & 0.12$\pm$0.007 & 1.128$\pm$0.058 & 0.10$\pm$0.022  & 1.066$\pm$0.014 & 0.13$\pm$0.008 & 0.624$\pm$0.023 & 1.136$\pm$0.04 & 0.10$\pm$0.016 & 0.394$\pm$0.127 \\
\bottomrule
\end{tabular}
\end{table*}

\subsection{Blue tilt}
\label{sec:tilt}

As mentioned in Section \ref{sec:color}, the blue GC 
candidates of NGC\,6861 seem to follow a colour-magnitude trend, known 
as `blue-tilt'. A similar phenomenon was observed in several elliptical 
galaxies (see \citealt{fort07,wehn08,harr09,peng08}). This trend has 
been interpreted as an increase of the metallicity with the GC mass, i.e., 
a mass-metallicity relation (MMR). Physics explanations for the 
`blue-tilt' were proposed by \citet{stra08} and \citet{bail09}, based 
on self-enrichment during a cluster's formation stage, where star formation is influenced by supernova feedback within the protocluster. 

\begin{figure}
\resizebox{0.85\hsize}{!}{\includegraphics{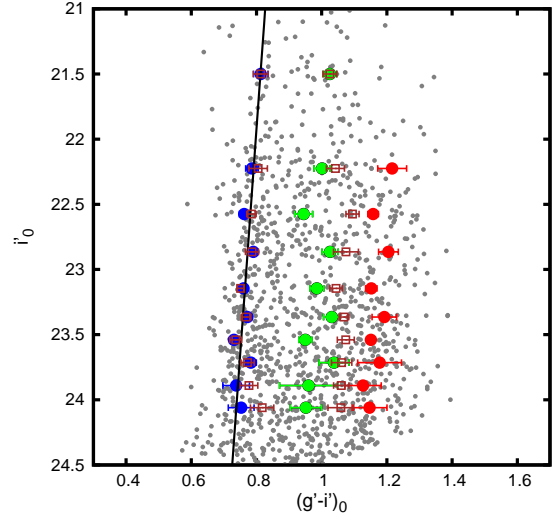}}
\caption{\textbf{Blue-tilt present in NGC\,6861 (black line) for trimodal case. Blue, green and red filled circles represent peak values in $(g'$--$i')_0$ colour obtained by RMIX for blue, green and red subpopulations, respectively. The brown open squares correspond to the peak values considering the bimodal case for blue and red subpopulations.}}
\label{fig11}
\end{figure}

With the aim to characterize this MMR in NGC\,6861, the sample of GC 
candidates was split in several magnitude bins of 100 objects. Subsequently, 
RMIX was used to obtain the locus of peaks in each bin for the three 
subpopulations {\bf{mentioned  in section \ref{sec:histo}}} (GMM gave us similar results, but RMIX was preferred 
due to its graphical direct interface). {\bf{In this case,}} we made 
homosedastic runs in order to facilitate the convergence (the three modes have the same variance) and tested 
with bins of different wides. 
In Figure\,\ref{fig11} we show, as an example,
 the mean values for blue, green and red  candidates obtained 
considering bins of 0.075 mag in an $i'$ versus $(g'$--$i')_0$ CMD. The 
figure shows that, as was suspected earlier from the colour-magnitude 
diagram, the blue peak indeed becomes redder at higher luminosities.  

\citet{harr09} found evidence that the blue-tilt in M87 begins to be detectable 
from magnitudes brighter than $M_I$$=$$-9.5$, exhibiting a nonlinear behaviour. Using the equation (1) 
from \citet{faif11} and the adopted distance to NGC\,6861, we 
obtained the equivalent value $i'_0\sim 23.24$ mag. Regarding the
slope of the tilt obtained by \citeauthor{harr09}, 
it was $d(g-i)/di=-0.021\pm0.004$. However, values slightly larger than
that were obtained for other galaxies. For example, \citet{faif11}
found  $d(g'-i')/di'=-0.032\pm0.003$ (for luminosity $M_I$$\leqslant$$-9.5$) for the giant elliptical galaxy M\,60, and \citet{wehn08} found $d(g'-i')/di'=-0.044\pm0.011$ (for luminosity $M_I$$\leqslant$$-9$) for NGC\,3311.

In order to compare with the different values 
from the literature, we made a linear least-squares fit to the 
colour as a function of $g'_0$ and $i'_0$ magnitudes. We use this
functional form as a first order approach and because 
it is worthless to use a more complex functional form for 
this particular sample of points.

As an example, we list here the fit obtained considering all 
the measured points, that obtained rejecting the points fainter 
than $i'_0=23.3$ mag ($M_I \sim -$9.5 mag), and that obtained 
rejecting only the brightest point, respectively:

\begin{equation}\label{eq:blue_tilt_all}
(g'-i')_0 = -0.026 (\pm0.007) i'_0 + 1.385(\pm0.163)
\end{equation}
\begin{equation}\label{eq:red_tilt_bright}
(g'-i')_0 = -0.025 (\pm0.011) i'_0 + 1.358(\pm0.268)
\end{equation}
\begin{equation}\label{eq:red_tilt_mbright}
(g'-i')_0 = -0.025 (\pm0.009) i'_0 + 1.344(\pm0.232)
\end{equation}

\noindent Although the behaviour of the data at the faint end is not so clear, 
we can see that the slopes for the different fits are identical. Therefore, we 
will adopt as the slope of the `blue tilt' in NGC\,6861, the first 
value, i. e., $d(g'-i')/di'$$=$$-0.026(\pm0.007)$.

{\bf {In order to facilitate the comparison of our results with 
other values present in the literature, we additionally run RMIX 
forcing it to fit a bimodal distribution to the same sample used 
in the trimodal case. In this opportunity, we made heteroscedastic 
runs (that means that different mode variances are allowed) and the 
resulting peaks of the two Gaussians for different magnitude bins are 
also shown in Figure \ref{fig11}. We can see that the blue peak is clearly 
recovered, and that most of them are very close to the points obtained in 
the trimodal case. On the other hand, the new red peaks are 
located between the ``green'' and ``red'' candidates of 
the tri-modal case. A linear fit to the blue peak points using 
the whole sample gives $(g'-i')_0=-0.021(\pm0.011) i'_0+1.258 (\pm0.257)$. 
When only the points brighter than $i'_0=23.3$ are included, the fit 
results in $(g'-i')_0=-0.033 (\pm0.010) i'_0+1.539(\pm0.239)$. Within 
the errors, these values are the same as those in equations (\ref{eq:blue_tilt_all})-(\ref{eq:red_tilt_mbright}) 
and very similar to those obtained for several elliptical galaxies, as
mentioned before.}}

{\bf{ The adopted values for the blue-tilt in the bimodal or 
trimodal cases, do }}not necessary imply that the MMR is 
linear for NGC\,6861. 
Our tests indicate that the slope, considering different 
sub-samples, seems to decrease as we consider the weakest 
magnitude points in the CMD (as found in other galaxies 
like those presented by \citealt{mies10} and \citealt{harr09}). 
However, the errors associated with those estimations are too big, 
indicating that we probably need a more deeper sample to obtain 
conclusive results. For that reason, in this paper we decided 
to compare our results only with estimates of linear fits in the 
literature.

As mentioned before, the different estimates found in the 
literature on the blue-tilt, mostly for elliptical 
galaxies, are similar to that obtained here. This is true even when
we compare slopes from other combinations of photometric filters.
For example, the results of \citet{ushe13} for NGC\,4278, or 
those from \citet{mies06b} and \citet{stra06} for M49, 
M60 and M87, are all around  $d(g'-z')/dz$$\sim$$-0.040$. This value is 
equivalent to that obtained by us in NGC\,6861 when we use 
the relation between the ACS $(g'$--$z')$ and $(g'$--$i')$ filters
from \citet{fort13}.

In a similar way as it was done for blue GCs, we 
performed linear fits for green and red candidates {\bf {in the trimodal case}}, obtaining 
$d(g'-i')/di'$$=$$-0.009(\pm0.022)$ and 
$d(g'-i')/di'$$=$$-0.019(\pm0.018)$, respectively. As it can be seen, the 
slope for green GCs is not significant whereas in the case of the red GCs, 
a weak red-tilt is obtained, though the value of this slope has a 
marginal significance.{\bf {When we force the bimodal case and we 
fit the read peak, we get $d(g'-i')/di'=0.012(\pm0.009)$. This slope has 
the opposite sign to that of the trimodal case. However, this marginally 
significant slope disappear when the brightest point is rejected. 
Therefore, we think that no slope is present in the red peak sample
when bimodality is assumed.}}

\subsection{Spatial distribution.}
\label{sec:dens}

\begin{figure}
\resizebox{0.85\hsize}{!}{\includegraphics{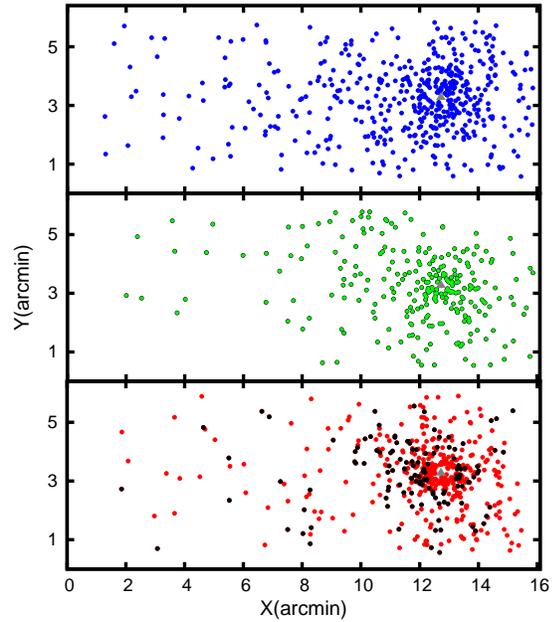}}
\caption{Projected spatial distribution of GC candidates. From top to 
bottom, the blue, intermediate and red subpopulations, according to the GMM 
separation. Black filled circles are the GC candidates 
with $(g'$--$i')$$>$1.16 which show an asymmetric distribution 
around the centre of NGC\,6861. The orientation is the same as 
Figure \ref{fig1}}
\label{fig12a}
\end{figure}

In Figure\,\ref{fig12a} we show the projected spatial distribution of the 
three subpopulations according to the separation of GMM mentioned in Section \ref{sec:histo}. 
It is clear in this figure the different spatial 
distributions of blue and red candidates. The latter looks more 
concentrated and flattened, resembling the starlight distribution of 
NGC\,6861 (see Subsection \ref{sec:azim}).
On the other hand, we have obtained the radial unidimensional distributions 
of GC candidates. The surface density for each radial bin was 
obtained counting the GC candidates in annuli, and applying a factor
to take into account the fraction of area effectively observed of each annulus.
In every case, the associated density error is given by Poisson statistics. 

\begin{figure}
\resizebox{0.85\hsize}{!}{\includegraphics{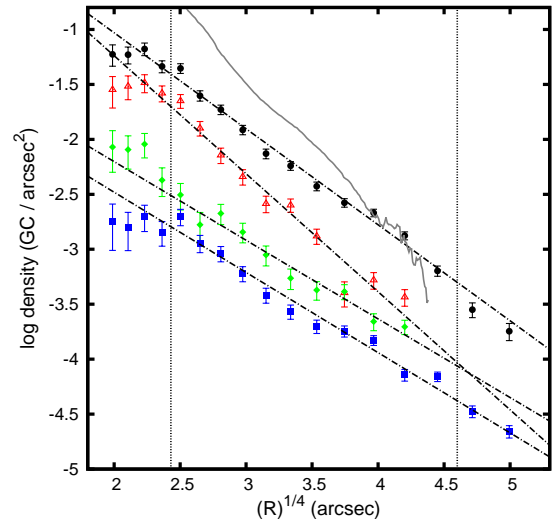}}
\caption{Projected density profile for red (triangles), green 
(diamond) and blue (squares) subpopulations, and all GC 
candidates (circles). Solid grey line 
shows the galaxy starlight profile, which has a similar slope to 
red clusters. All profiles were fitted with a power-law and a de 
Vaucouleurs law. The vertical dotted lines indicate the range 
used to make the fits (2.4$<r^{1/4}<$4.6 arcsec). The profiles were 
shifted to avoid overlapping.}
\label{fig12}
\end{figure}

For the entire sample of candidates, we used concentric 
circular annuli with $\Delta$ $log\,r$ = 0.1, 
 corrected by background contamination, reaching $\sim$ 10 arcmin 
($\sim$100 kpc) from the galactic centre. 
Subsequently, we have obtained separated profiles for the 
three subpopulations: red, green and blue (Figure \ref{fig12}).
Due to the elongation showed by the red and green subpopulations, the 
profiles associated with them were obtained using concentric 
elliptical annuli, according to the position angle and ellipticity 
values obtained in Section \ref{sec:azim}. In the case of blue
GC candidates, we didn't detect any significant elongation, therefore 
we decided to obtain their density profile using circular annuli.   
It can be seen in Figure \ref{fig12}, that the 
red subpopulation presents a stronger spatial concentration 
towards the galaxy than the blue GCs. On the other hand, blue and green
subpopulations show very similar radial profiles.
Regarding the extension of the GC system of NGC\,6861, Figure \ref{fig12a} 
shows candidates at galactocentric distances even larger than 10 
arcmin ($\sim$100 kpc). Although at such distance our areal completeness 
is small, we can not rule out that some of them are bona fide GCs. 

In order to assess the slope of the profile, we fit a power law and a 
de Vaucouleurs law ($r^{1/4}$) to the whole sample, and to the red, 
green, and blue candidates, respectively. We present the obtained values 
in Table \ref{tab5}. These values are similar to 
those found in \citet{faif11} for NGC\,524 and NGC\,3115, other 
S0 galaxies (see their Table 5). It is interesting to note
that the green candidates show identical slopes than the blue ones.
However, Figure \ref{fig12} shows that green candidates display
a possible excess of objects at the inner region of the system,
not present in the blue or in the red profiles.

\begin{table}
\centering
\caption{Slope values for power law and de Vaucouleurs law, in the density 
profiles for the whole sample and for the blue, green and red subpopulations.}
\label{tab5}
\scriptsize
\begin{tabular}{ccc}
\toprule
\toprule
\multicolumn{1}{c}{\textbf{Population}} &
\multicolumn{1}{c}{\textbf{Power law}}  &
\multicolumn{1}{c}{\textbf{de Vaucouleurs}} \\
\midrule
All     & -1.60$\pm$0.04  & -0.87$\pm$0.03 \\
Blue    & -1.47$\pm$0.07 & -0.73$\pm$0.04 \\
Green   & -1.35$\pm$0.08 & -0.71$\pm$0.05 \\
Red     & -2.04$\pm$0.11 & -1.07$\pm$0.08 \\
\bottomrule
\end{tabular}
\end{table}

In summary, from Figures \ref{fig12a} and \ref{fig12} we can see that 
the spatial distribution for the blue clusters is radially shallow and 
circularly symmetric; for the green candidates is shallow and elliptical; 
and the red GCs distribution is stepper and elliptical.

In order to see if there is any indication of sub-structure in the spatial 
distribution of our GC candidates, we analyzed the appearance of the xy 
plots by taking the candidates in different $(g'$--$i')$ colour windows. 
That exercise has shown that GC candidates with $(g'$--$i')$$>$1.16 present 
an asymmetric distribution around de centre of NGC\,6861. In the lower panel 
of Figure \ref{fig12a}, we plotted those candidates as black points. In 
order to quantify this asymmetry we have done an easy test: we have taken 
objects with $30<R_{gal}<160$ arcsec (complete areal coverage), and then 
counted them splitting the sample between the left (SE) and the right hand 
side of the galaxy (NW). We have got 69 and 35 candidates, respectively. 
Those numbers show that there are two times more such candidates to the 
SE of the galaxy centre than to the NW. This could be not a contamination 
effect because, according to the analysis of our comparison 
field, we can expect few contaminant object in that area ($<$10). 
Inhomogeneities in GCs spatial distributions were reported in other 
early type galaxies like NGC\,4261, NGC\,4649 and NGC\,4278 
\citep{dabr13,dabr13b,dabr14}. These features, like fossil remnants, are 
strongly suggestive of the accretion and/or merger of some lower mass neighbors.

\subsection{Azimuthal properties}
\label{sec:azim}

\begin{figure}
\resizebox{0.9\hsize}{!}{\includegraphics{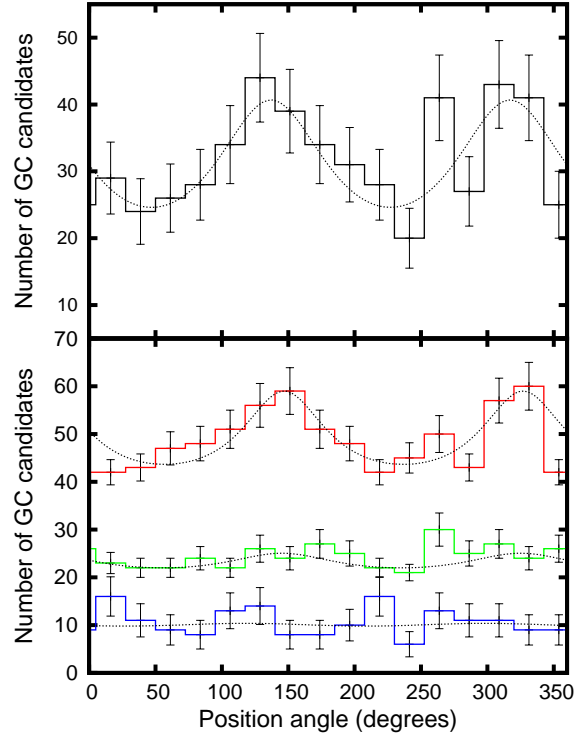}}
\caption{Azimuthal distribution for the entire sample of GC candidates 
(upper panel), and red, green and blue subpopulations (bottom panel), 
within a ring around the galaxy as described in 
the text. In order to avoid overlapping, the curves corresponding 
to green and red candidates were vertically shifted adding 18 
and 35 to the counts.}
\label{fig13}
\end{figure}

In Subsection \ref{sec:dens}, Figure \ref{fig12a}, we saw that red candidates
show a flattened projected spatial distribution. In order to check
that, we use the expression of \citet{mcla94}: 

\begin{equation}\label{eq:azim}
\sigma(R,\theta)=kR^{-\alpha}[cos^2(\theta-PA)+(1-e^2)^{-2}sin^2(\theta-PA)]^{-\alpha/2}
\end{equation}

\noindent  to study the GC azimuthal distribution, where $\sigma$ is the 
number of GC candidates. We consider the whole sample, and the three 
subpopulations of GC candidates. We fit
the position angle ($PA$), measured counterclockwise from the north, 
and the ellipticity ({\it e}) in Equation \ref{eq:azim}. 
There, $\alpha$ is the value of the power-law exponent previously obtained 
in the surface density fit (see Section \ref{sec:dens}). In order to bring 
together the largest number of GC candidates, and avoiding correction 
for areal incompleteness, we took a circular ring around the galaxy 
with 31 $<$$R_{gal}$$<$ 101 arcsec. Subsequently, we divided 
the ring in wedges of 22.5\degree, and we fit the 
Equation \ref{eq:azim}, allowing the normalization constant ($k$),   
the $PA$ and the ellipticity to vary. In Figure \ref{fig13}, we show the
histograms of the azimuthal distributions for the whole sample, 
for the blue, green and red GC candidates and the obtained fits. 
In Table \ref{Tab6}, we present 
the values obtained for the two fitted parameters. There we also list 
the results for the X-ray emission \citep{mach10}.

Figure \ref{fig13} shows that the spatial distribution of the whole 
sample of GC candidates, exhibits strong signs of elongation. The 
$PA$ results identical within the error to that of the galaxy inside a 
semi-major axis of 120 arcsec (see Subsection \ref{sec:elipse}). 
As far as ellipticity is concerned, the values obtained for
green, red and all GC candidates are slightly bigger than that of the
galaxy. In particular, red candidates show a significantly higher ellipticity
than that of the starlight. It is interesting to notice that,  
according to Figure \ref{fig13}, the behaviour of the GC candidates is 
led by the red and green ones.
For blue candidates we didn't find any significant elongation 
and, then, the $PA$ results unconstrained.

\begin{table}
\centering
\caption{Values of position angle and ellipticity for the three 
subpopulations and for the whole sample of GC candidates. In addition, 
we include the results for X-ray measurements from \citet{mach10}.}
\label{Tab6}
\scriptsize

\begin{tabular}{lp{0.4cm}p{0.9cm}p{0.8cm}p{0.8cm}cll}
\toprule
\toprule
\multicolumn{1}{c}{\textbf{Param.}} &
\multicolumn{1}{c}{\textbf{Blue}} &
\multicolumn{1}{c}{\textbf{Green}} &
\multicolumn{1}{c}{\textbf{Red}} &
\multicolumn{1}{c}{\textbf{All}} &
\multicolumn{2}{c}{\textbf{X}} \\
\multicolumn{5}{c}{} &
\multicolumn{1}{c}{{\it r\,$\leqslant$7.3 kpc}} &
\multicolumn{1}{c}{{\it r\,$\leqslant$42 kpc}} \\
\midrule
$PA$ & - & 145\degree$\pm$13\degree & 146.8\degree$\pm$5\degree & 137\degree$\pm$6\degree &  128\degree$\pm$7\degree & - \\
$e$ & - & 0.54$\pm$0.1 & 0.64$\pm$0.04 & 0.52$\pm$0.05 & 0.40$\pm$0.04 & 0.19$\pm$0.05 \\
\bottomrule
\end{tabular}
\end{table}

\subsection{Comparison with the hot gas emission}

In their study of the GC systems of the giant elliptical galaxy NGC\,1399, 
\citet{fort05} have shown that there is a similarity between 
the projected density profile of the 
blue GC subpopulation and that of the X-ray surface brightness profile of that
galaxy. More recently, \citet{forb12} have extended the study to other
eight giant elliptical galaxies and they have found that
the same behaviour is present in their sample. Therefore, as far as 
we are aware, these nine galaxies are the only cases where a 
direct comparison between the spatial distribution of the blue GCs 
and the X-ray emission has been made and published. Excluding NGC\,720, all of 
them are massive and giant ellipticals in groups and clusters. 

\citet{fort07,fort09,fort14} suggested that the red GCs 
are associated with the bulge stellar component in early type 
galaxies, while the blue GCs would be linked to the halo 
component. In this context, it is natural
to think that blue GCs and hot gas can share the same gravitational 
potential in equilibrium \citep{forb12}. However, it is important 
to notice that similar projected 2-d distributions do not 
necessarily imply similar 3-d spatial distribution.

Thus, we decided to make a comparison between the slope of the blue GCs
density profile and that of the emitting hot gas. Fortunately, 
there is a complete analisys of the X-ray emission around NGC\,6861 and
its neighbour NGC\,6868 in the Telescopium galaxy group presented by
\citet{mach10}. For both galaxies \citeauthor{mach10} quantify the 
X-ray emission through double $\beta$-model profiles.
The inner component could be associated with the galaxy emission and
the outer one with a subgroup hot halo. In this context \citeauthor{mach10} 
suggested that NGC\,6868 and NGC\,6861 may each be the dominant
galaxy in a merging galaxy subgroup.
The $\beta$-model is commonly used to reproduce the observed 
X-ray surface brightness profiles in galaxies \citep{osul03}, and is 
defined as:

\begin{equation}\label{xprof}
S(r) = S_0[1+(r/r_c)^2]^{-3\beta_x+0.5}
\end{equation}

\noindent The relevant parameters in the model are the core radius 
($r_c$) and the slope profile ($\beta$$_x$), this latter defines the behaviour 
of the density profile for $r$$>$$r_c$.
In the case of NGC\,6861, \citeauthor{mach10} found the following 
$\beta$-model parameters: $r_c$=0.5(11.9) kpc and $\beta$=0.61(0.38) 
for the inner (outer) $\beta$-model components, respectively. 
 
\begin{figure}
\resizebox{0.85\hsize}{!}{\includegraphics{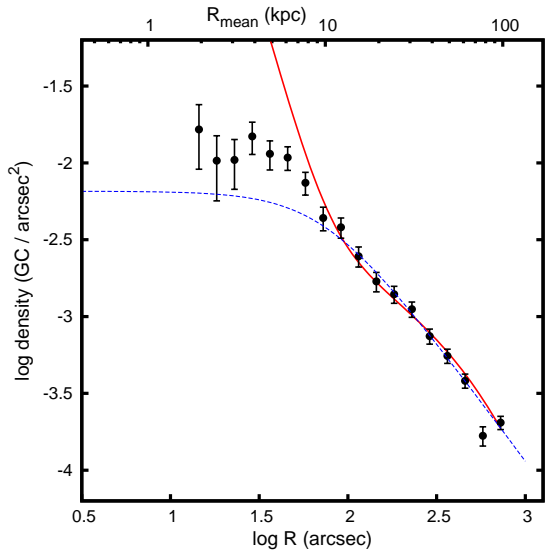}}
\caption{Globular cluster density profile for the blue subpopulation 
(filled black circles) and surface brightness profile in X-ray (red line) 
of NGC\,6861 (Jones et al. in prep). The profiles were shifted on the 
vertical axis for a better comparison. Blue dashed line shows the 
single $\beta$-model fitted to the GCs density. The agreement between the 
blue GCs and X-ray profiles outside 10 kpc is remarkable.}
\label{fig12b}
\end{figure}

In order to quantitatively compare the slopes of blue GCs and 
X-ray profiles, we have counted blue cluster candidates
in concentric circular annuli. Figure \ref{fig12b} shows
counts as small filled circles and the corresponding 
Poissonian error bars. As a red solid line we plotted the double
$\beta$-model from Jones et al. (in preparation), shifted vertically
to have a good match. As it can be seen, the X-ray profile
shows an excellent agreement with the blue globulars density 
for $r$$>$50 arcsec ($>$ 10 kpc). Subsequently, we have fitted 
the density profile considering points in the range 10.6 $<$$r$$<$ 67.1 kpc by 
employing a single $\beta$-model.
The obtained value of the slope is $\beta\sim0.38-0.44$,
which is very similar to that of \citeauthor{mach10}
However, due to the restricted radial range included in the fit,
the $r_c$ parameter results not well constrained. In order to
obtain a more reliable $\beta$ value, we have 
fitted again the data, but keeping fixed the $r_c$ parameter at the value 
given by \citeauthor{mach10} We have obtained $\beta=0.42 \pm 0.01$.  
We plotted this fit as a dashed blue line in Figure \ref{fig12b}.
Again, the agreement is excellent showing that
outer blue GCs present a similar projected distribution 
to that of the sub-group X-ray component.

\subsection{Luminosity Function} 
\label{sec:lumin}

The adopted distance of 28.1 Mpc for NGC\,6861 in this work, translates in
a distance modulus of $(m-M)=32.24$ mag. Therefore, taking an approximated
turn over of $M_V$$=$$-$7.5 mag. (e.g. \citealt{harr01}), and equation 2 from
\citet{faif11}, we estimated a value for the peak
of the GC luminosity function (GCLF) in the $g'$ band at $\sim$25.1 mag.
As we showed in Subsection \ref{sec:test}, the completeness
of our photometry is greater than 80 per cent for $g'_0<$25.5 mag.
This indicates that we will be able to obtain
a good representation of the integrated GCLF for NGC\,6861. 

\begin{figure}
\resizebox{0.85\hsize}{!}{\includegraphics{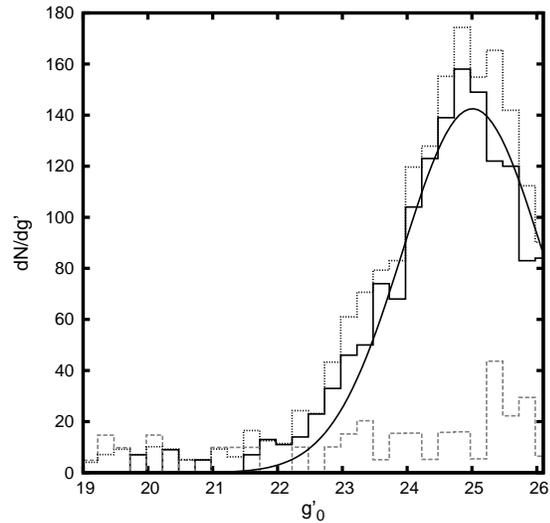}}
\caption{The dotted histogram shows counts of globular cluster candidates 
considering bins of 0.25 mag. After applying completeness and background corrections, we obtain the solid line histogram. The adopted normalized background is shown as a dashed histogram. The solid line represents the Gaussian fit to the corrected globular cluster luminosity function.
}
\label{fig14}
\end{figure}

In order to do that, we counted candidates in bins 
of 0.25 mag. This raw luminosity distribution is
shown in Figure \ref{fig14} as a dotted line histogram. 
Subsequently we have corrected the counts for completeness 
effects and background contamination and plotted it as 
solid histograms in Figure \ref{fig14}. This figure shows that
a peak around $g'_0\sim 25$ mag is clearly detected. 

Although different options exist in the literature to model 
the GCLF (see \citealt{jord07}), 
we have adopted here a Gaussian form. The main reason behind this 
election is that there is a historic use of Gaussian representation 
as distance indicator (\citealt{jaco92,vill10}).
Therefore, we fit a Gaussian function to our histogram, and 
determined the position of the turn-over magnitude ($TO$), 
and the dispersion of the GCLF. In the fit we included all the
bins with magnitude $g'$$<$26.2 mag. The values obtained for 
both parameters are listed in Table \ref{Tab7}, and 
Figure \ref{fig14} shows the obtained fit.
The value of $\sigma_g'$ is in very good agreement 
with the relation between the intrinsic Gaussian dispersion of the 
g-band GCLF, and the absolute blue magnitude of the parent 
galaxy, $\sigma_{g}$--$M_B$, presented in the figure 9 of \citet{jord07}.

\begin{figure}
\resizebox{0.85\hsize}{!}{\includegraphics{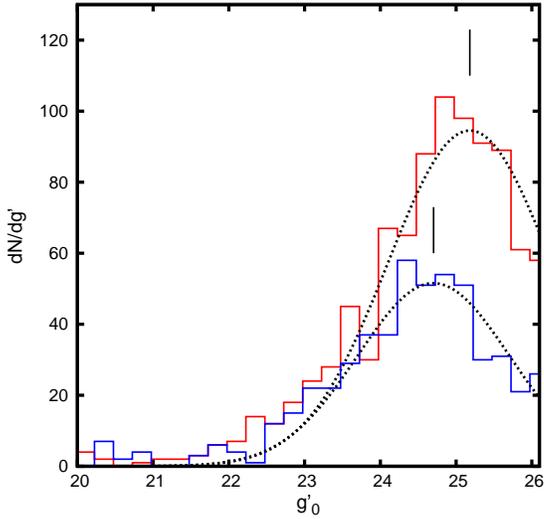}}
\caption{Corrected luminosity functions for blue and red subpopulations. Dotted lines and vertical short lines represent the Gaussian fit and $TO$ magnitude, respectively.}
\label{fig15}
\end{figure}

Our next step was to obtain the GCLF for the different subpopulations
detected in NGC\,6861. Figure \ref{fig15} shows the blue and red corrected histograms and the fitted Gaussian models. As we did not detect any difference between the fit for the green and red candidates, we considered both subsamples together in the fit. The values of the obtained parameters are
listed in Table \ref{Tab7}.

It is interesting to notice that we have obtained a significant 
difference between the turn over magnitudes of blue and red clusters.
We can see from Table \ref{Tab7} that $TO_{red}-TO_{blue}\sim 0.48$ mag.
It is immediate to think in the existence of metallicity effects
behind this result. \citet{ashm95} have investigated the shift
in the peak of the GCLF as a function of metallicity for 
the $B$, $V$, $R$, $I$ and $J$ filters. They based their 
results on simulations that included the cluster masses 
drawn from a parent population similar to that of the MW, and a 
mean metallicity ranging from -0.2 dex to -1.6 dex. According 
to their table 3, and assuming mean $[Fe/H]$ of -1.6 and
-0.6 dex for the blue metal poor and 
red metal rich GCs, respectively, we can expect a turn over 
shift of around 0.3 mag in the $V$ band. Red GCs in NGC\,6861 are probably 
more metal rich than those of the MW. Therefore, the turn over shift 
in this S0 galaxy could be larger than in our galaxy.
Unfortunately, \citet{ashm95} did not include the
Sloan filters in their calculations. For that reason we 
transformed the turn over magnitudes for blue and red GCs
using equation 2 from \citet{faif11}, and adopting as
mean colours 0.78 and 1.13 mag for blue and red candidates, respectively.
This translates in a 0.43 mag shift for the turn over magnitude 
in the $V$ band. If we assume that blue GCs in NGC\,6861 are similar to those
of the MW, according to figure 7 of \citet{rejk12}, the mean 
metallicity of red GC in this galaxy should be $\sim$ -0.25 dex.

This result shows that it may not be a good idea to measure distances
using the GCLF obtained in blue bands for the whole sample. It is possible to 
minimize the metallicity effect by observing in 
the near IR regime (like $z$, \citealt{vill10}). However, 
another option is to use the turn over magnitude of the blue GCs
as a distance indicator. \citet{dicr06} have shown that
adopting a common LMC fiducial distance scale of $(m-M)=18.50$ mag, 
the $TO$ magnitudes of metal-poor GCLF in the MW, M31 and the more
distant galaxy sample of \citet{lars01} are in perfect 
mutual agreement with $M_V$$=$$-$7.66 mag.

Taking \citet{dicr06} turn over magnitude for 
the blue GCs, the corresponding mean $(g'$--$i')$ colours from Table \ref{Tab4}
 and using equation 2 from \citet{faif11}, we got 
$M_{g'}$$=$$-$7.39 mag. With this value and the apparent magnitude for
the turn over of the blue GCLF from Table \ref{Tab7}, we got a distance 
modulus of $(m-M)=32.1$ mag. This value is in very good agreement with that
of \citet{tonr01}, $(m-M)=32.24\pm0.36$. mag.
 
\begin{table}
\centering
\caption{Parameters obtained by fitting Gaussian functions to the $g'$-band
GCLF. Turn-over magnitude ($TO$) and dispersion ($\sigma$) for the whole 
sample, and blue and red subpopulations, are shown.}
\label{Tab7}
\scriptsize
\begin{tabular}{ccc}
\toprule
\toprule
\multicolumn{1}{c}{\textbf{Population}} &
\multicolumn{1}{c}{\textbf{$TO$}} &
\multicolumn{1}{c}{\textbf{$\sigma$}} \\
\midrule
All     & 25.00$\pm$0.05  & 1.08$\pm$0.06 \\
Blue    & 24.70$\pm$0.06  & 1.00$\pm$0.06 \\
Red     & 25.18$\pm$0.08  & 1.07$\pm$0.08 \\
\bottomrule
\end{tabular}
\end{table}


\subsection{Total number of GCs and Specific Frequency}

In Section \ref{sec:dens} we determined the density profiles for 
the different subpopulations and for the whole system. By extrapolating 
the density profile from $\sim$30 to 600 arcsec, we got 1834 GC.
This last value is equivalent to $\sim$100 kpc at the adopted distance
of NGC\,6861. We have taken this external cut because 
the analysis presented in that subsection showed that 
there are  GCs even at such a large galactocentric distance. Besides, that 
value is in agreement with several wide field studies of different GC 
systems of massive galaxies (\citealt{bass06,harr09b}). 
In the inner region ($\lesssim$30 arcsec), we assume that the 
density of GCs is constant. This gives us 160 more GCs.

These 1994 GCs, according to the $TO$ value obtained for the whole
sample of candidates, correspond 
to 67 per cent of GCs brighter than $g'_0$$=$25.5 mag, leaving 33 per 
cent fainter. Considering a $\pm20$ per cent variation in the extension
of the surface density profile due to the uncertainties in their parameters,
by integrating the LF profile over all magnitudes, we estimate a total number 
of $3000\pm300$ GCs.

Using the apparent visual magnitude $V_T^0$$=$10.97 (NED; an 
identical value was published by \citealt{ho11}) and the distance 
module $(m$--$M)$$=$32.24 for NGC\,6861, we obtain as its 
absolute magnitude $M_V$$=$$-$21.27. From this value and from the total 
number of GCs, we can calculate the specific frequency ($S_N$), defined 
by \citet{harr81} (Equation \ref{specific}).
 
\begin{equation}\label{specific}
S_N = N_{GC}10^{0.4(M_V+15)}.
\end{equation}

Therefore, for NGC\,6861 we obtain a surprisingly high value of 
$S_N=9.2\pm2.2$. On the other hand, if we adopt the distance modulus obtained 
from the $TO$ magnitude for blue GCs in previous section, we 
get $M_V$$=$$-$21.12 mag, and a $S_N=10.6\pm2.1$. These are extremely 
high values of $S_N$, similar to that of giant cluster dominant early 
type galaxies like M87. \citet{harr13} presented a compilation of global 
properties of GC systems. We can see that there are only two S0 galaxies 
with $D<100$ Mpc and $M_V$$<$$-$20 mag which present $S_N$$>$7 
in that sample. They are NGC\,1400 and IC\,3651. However, checking the 
source of that values, we can see that in the case of NGC\,1400, 
\citet{perr97} found $N_t=922\pm280$ GCs, and a $S_N=5.2\pm2$. On the 
other hand, \citet{mari02} list a value of $S_N=5.7\pm1.8$ for IC\,3651. 
However, this relies on an indirect estimation of $M_V$ for that 
galaxy as it does not appear in the RC3 catalog, and on an indirect 
estimation of $N_t$ using the Surface Brightness Fluctuation method. In this 
context, NGC\,6861 seems to be a more definite case of a high $S_N$ lenticular 
galaxy.

High $S_N$ values are usually found in low and high mass early-type galaxies. 
This is evident in the ``U'' shape plot of $S_N$ vs $M_V$ like that presented in 
Fig. 10 in \citet{harr13}. As $S_N$ measures the formation efficiency 
of GCs relative to field stars, high $S_N$ galaxies can be understood as 
``field-stars deficient'' or ``cluster rich'' systems. However, several 
studies provided growing evidence in favour of the first situation. 
Specifically, \citet{mcla99} has shown that the enhanced 
$S_N$ value for the most massive galaxies in his sample can be accounted 
for if the mass in GCs is normalized to the total baryonic mass of 
the host (i.e. X-ray emitting gas and the stellar component), 
$\epsilon=M_{GCS}/M_b\sim0.0026$. In this context, according its $S_N$ 
value, NGC\,6861 could have a massive baryonic halo associated with the 
hot gas emitting in X-ray.\ 

Recently, a more fundamental ratio was defined: the ratio of the mass in GCs 
to the total galaxy halo mass, $\eta=M_{GCs}/M_h$. This ratio seems to be 
essentially constant over the whole mass range of galaxies 
(\citealt{geor10}; \citealt{harr13}). 
Using a comprehensive database of 307 galaxies, \citet{huds14} 
confirmed this suggestion through a new calibration of galaxy halo masses 
entirely based on weak lensing. They obtained a value 
of $<$$\eta$$>=$3.9 $\pm$ 0.9$\times$ $10^{-5}$. Following Harris et al. (2013) we 
estimated the total mass enclosed in the GCS of NGC\,6861 
(adopting a $(M/L)_V$$=$2), $M_{GCs}$$=$9 $\pm$ 1$\times$$10^8$ $M_\odot$, and 
therefore we used the  $<\eta>$ value to estimate the total mass 
of the halo of NGC\,6861. We found $M_h$$=$2.3 $\pm$ 0.5$\times$$10^{13}$$M_\odot$. 
This value indicates again that according its GC system, we can 
expect a massive halo in NGC\,6861.



\section{CONCLUSIONS}

Using deep images taken with GEMINI/GMOS, we obtained 
multi-colour photometry of three fields around NGC\,6861. This 
galaxy presents unique features for an S0 type, summarized as follows:

\begin{itemize}

\item NGC\,6861 presents signs of past interaction or merger, like the 
non-concentric isophotes shown by the galaxy. In addition, we 
have found several different structures that become visible when 
the ratio between the original image and the smooth 
ellipse model is analyzed.

\item The galaxy has a large GC system which extends up to 100 kpc. 
By analyzing the colour-colour and colour-magnitude diagrams, we 
detected at least two subpopulations of GC, blue and red, with 
colour peaks at $(g'$--$i')_0\sim0.79$ and $(g'$--$i')_0\sim1.07$ mag, 
respectively. By examining the colour histograms and the CMD, we 
inferred the possible existence of a third subpopulation with 
intermediate colours ($(g'-i')_0 \sim0.95$).
We have tentatively separated the cluster candidates in blue, green 
and red subpopulations. However, spectroscopic and/or IR data are 
needed to perform an age-metallicity 
study, and confirm or rule out the existence of this third subpopulation.

\item The presence of a blue-tilt is detected in the blue subpopulation, 
probably being the first case in this galaxy type.

\item Regarding the spatial distribution of the blue and green subpopulations, 
their density profiles show similar slopes.

\item Outside 10 kpc, the blue subpopulation shows a remarkable 
similarity between the projected density profile and the X-ray surface 
brightness profile of the galaxy. This result would indicate 
that NGC\,6861 could be the dominant galaxy in a 
galaxy sub-group, as mentioned by \citet{mach10}.

\item The azimuthal distribution of the red and green GC candidates 
exhibit strong signs of elongation, with an ellipticity slightly 
larger than that of the galaxy.

\item By comparing the galaxy light with the red subpopulation, we 
can see that in the inner region they both share similar features 
such as $PA$, colour and slope of the density profile.

\item Red GC candidates with colours $(g'$--$i')_0$$>$1.16 show an 
asymmetric spatial distribution around the galaxy. Similar features 
were reported in other galaxies, 
suggesting accretion and/or merger of a lower mass neighbour.

\item We obtained the global GCLF of NGC\,6861, as well as for the 
blue and red subpopulations separately.
A significant difference between the $TO$ values of both subpopulations 
in the $g'$ band, was obtained ($TO_{red}-TO_{blue}\sim 0.48$ mag). The 
existence of metallicity effects might be behind this result.
 
\item Using the blue GC candidates, we estimated a new distance modulus 
($m-M=32.1$ mag), which shows a good agreement with that of \citet{tonr01}.

\item We have estimated a total population of $3000\pm300$ GCs for NGC\,6861.

\item From the distance given by \citet{tonr01}, we obtained a high 
specific frequency of $S_N=9.2\pm2.2$, and from our own distance 
estimation, a higher value of $S_N=10.6\pm2.1$. 
As far as we know, this is the highest $S_N$ measured for a 
lenticular galaxy up to date. This high $S_N$ value 
could be interpreted as a sign of the presence of a massive halo in NGC\,6861.
 
\end{itemize}

\section*{Acknowledgements}
We warmly thank M. Machacek for providing the X-ray surface brightness 
profile of NGC\,6861, and Sergio Cellone and Analia Smith Castelli for 
reading of the manuscript and several useful comments.
This work was funded with grants from Consejo Nacional de Investigaciones
Cientificas y Tecnicas de la Republica Argentina, and Universidad Nacional
de La Plata (Argentina). Based on observations obtained at the Gemini Observatory, 
which is operated by the Association of Universities for Research in Astronomy, Inc., under a cooperative agreement with the NSF on behalf of the Gemini partnership: the National 
Science Foundation (United States), the National Research Council (Canada), 
CONICYT (Chile), the Australian Research Council (Australia), Minist\'{e}rio da 
Ci\^{e}ncia, Tecnologia e Inova\c{c}\~{a}o (Brazil) and Ministerio de Ciencia, 
Tecnolog\'{i}a e Innovaci\'{o}n Productiva (Argentina). 
The Gemini program ID are GS-2010B-Q-2 and GS-2011A-Q-81. This research has made use of the 
NED, which is operated by the Jet Propulsion Laboratory, Caltech, under contract with the 
National Aeronautics and Space Administration. 

\bibliographystyle{mn2e}
\bibliography{bibliography}

\begin{thebibliography}{}

\bibitem[\protect\citeauthoryear{{Arag\'on-Salamanca}, {Bedregal} \&
  {Merrifield}}{{Arag\'on-Salamanca} et~al.}{2006}]{arag06}
{Arag\'on-Salamanca} A.,  {Bedregal} A.~G.,    {Merrifield} M.~R.,  2006, A\&A,
  458, 101

\bibitem[\protect\citeauthoryear{{Ashman}, {Conti} \& {Zepf}}{{Ashman}
  et~al.}{1995}]{ashm95}
{Ashman} K.~M.,  {Conti} A.,    {Zepf} S.~E.,  1995, AJ, 110, 1164

\bibitem[\protect\citeauthoryear{{Ashman} \& {Zepf}}{{Ashman} \&
  {Zepf}}{1992}]{ashm92}
{Ashman} K.~M.,  {Zepf} S.~E.,  1992, ApJ, 384, 50

\bibitem[\protect\citeauthoryear{{Bailin} \& {Harris}}{{Bailin} \&
  {Harris}}{2009}]{bail09}
{Bailin} J.,  {Harris} W.,  2009, ApJ, 695, 1082

\bibitem[\protect\citeauthoryear{{Barr}, {Bedregal}, {Arag\'on-Salamanca},
  {Merrifield} \& {Bamford}}{{Barr} et~al.}{2007}]{barr07}
{Barr} J.~M.,  {Bedregal} A.~G.,  {Arag\'on-Salamanca} A.,  {Merrifield} M.~R.,
     {Bamford} S.~P.,  2007, A\&A, 470, 173

\bibitem[\protect\citeauthoryear{{Bassino}, {Faifer}, {Forte}, {Dirsch},
  {Richtler}, {Geisler} \& {Schuberth}}{{Bassino} et~al.}{2006}]{bass06}
{Bassino} L.~P.,  {Faifer} F.~R.,  {Forte} J.~C.,  {Dirsch} B.,  {Richtler} T.,
   {Geisler} D.,    {Schuberth} Y.,  2006, A\&A, 451, 789

\bibitem[\protect\citeauthoryear{{Beasley}, {Baugh}, {Forbes}, {Sharples} \&
  {Frenk}}{{Beasley} et~al.}{2002}]{beas02}
{Beasley} M.~A.,  {Baugh} C.~M.,  {Forbes} D.~A.,  {Sharples} R.~M.,    {Frenk}
  C.~S.,  2002, MNRAS, 333, 383

\bibitem[\protect\citeauthoryear{{Bekki}, {Couch} \& {Shioya}}{{Bekki}
  et~al.}{2002}]{bekk02}
{Bekki} K.,  {Couch} W.~J.,    {Shioya} Y.,  2002, ApJ, 577, 651

\bibitem[\protect\citeauthoryear{{Bertin} \& {Arnouts}}{{Bertin} \&
  {Arnouts}}{1996}]{bert96}
{Bertin} E.,  {Arnouts} S.,  1996, A\&AS, 117, 393

\bibitem[\protect\citeauthoryear{{Blom}, {Spitler} \& {Forbes}}{{Blom}
  et~al.}{2012}]{blom12}
{Blom} C.,  {Spitler} L.~R.,    {Forbes} D.~A.,  2012, MNRAS, 420, 37

\bibitem[\protect\citeauthoryear{{Brodie}, {Usher}, {Conroy}, {Strader},
  {Arnold}, {Forbes} \& {Romanowsky}}{{Brodie} et~al.}{2012}]{brod12}
{Brodie} J.~P.,  {Usher} C.,  {Conroy} C.,  {Strader} J.,  {Arnold} J.~A.,
  {Forbes} D.~A.,    {Romanowsky} A.~J.,  2012, ApJ, 759, 33

\bibitem[\protect\citeauthoryear{{Caso}, {Richtler}, {Bassino}, {Salinas},
  {Lane} \& {Romanowsky}}{{Caso} et~al.}{2013}]{caso13}
{Caso} J.~P.,  {Richtler} T.,  {Bassino} L.~P.,  {Salinas} R.,  {Lane} R.~R.,
   {Romanowsky} A.,  2013, A\&A, 555, 56

\bibitem[\protect\citeauthoryear{{Cen}}{{Cen}}{2001}]{cen01}
{Cen} R.,  2001, ApJ, 560, 592

\bibitem[\protect\citeauthoryear{{C\^ot\'e}, {Marzke} \& {West}}{{C\^ot\'e}
  et~al.}{1998}]{cote98}
{C\^ot\'e} P.,  {Marzke} R.~O.,    {West} M.~J.,  1998, ApJ, 501, 554

\bibitem[\protect\citeauthoryear{{Cowie} \& {Songalia}}{{Cowie} \&
  {Songalia}}{1977}]{cowi77}
{Cowie} L.,  {Songalia} A.,  1977, Nature, 266, 501

\bibitem[\protect\citeauthoryear{{D'Abrusco}, {Fabbiano} \&
  {Brassington}}{{D'Abrusco} et~al.}{2014}]{dabr14}
{D'Abrusco} R.,  {Fabbiano} G.,    {Brassington} N.~J.,  2014, ApJ, 783, 19

\bibitem[\protect\citeauthoryear{{D'Abrusco}, {Fabbiano}, {Mineo}, {Strader},
  {Fragos}, {Kim}, {Luo} \& {Zezas}}{{D'Abrusco} et~al.}{2013a}]{dabr13}
{D'Abrusco} R.,  {Fabbiano} G.,  {Mineo} S.,  {Strader} J.,  {Fragos} T.,
  {Kim} D.-W.,  {Luo} B.,    {Zezas} A.,  2013a, ApJ, 783, 18

\bibitem[\protect\citeauthoryear{{D'Abrusco}, {Fabbiano}, {Strader}, {Zezas},
  {Mineo}, {Fragos}, {Bonfini}, {Luo}, {Kim} \& {King}}{{D'Abrusco}
  et~al.}{2013b}]{dabr13b}
{D'Abrusco} R.,  {Fabbiano} G.,  {Strader} J.,  {Zezas} A.,  {Mineo} S.,
  {Fragos} T.,  {Bonfini} P.,  {Luo} B.,  {Kim} D.-W.,    {King} A.,  2013,
  ApJ, 773, 87

\bibitem[\protect\citeauthoryear{{de Vaucouleurs}, {de Vaucouleurs}, {Corwin},
  {Buta}, {Paturel} \& {Fouque}}{{de Vaucouleurs} et~al.}{1991}]{deva91}
{de Vaucouleurs} G.,  {de Vaucouleurs} A.,  {Corwin} H.~G.,  {Buta} R.~J.,
  {Paturel} G.,    {Fouque} P.,  1991, S\&T, 82, 621

\bibitem[\protect\citeauthoryear{{Di Criscienzo}, {Caputo}, {Marconi} \&
  {Musella}}{{Di Criscienzo} et~al.}{2006}]{dicr06}
{Di Criscienzo} M.,  {Caputo} F.,  {Marconi} M.,    {Musella} I.,  2006, MNRAS,
  365, 1357

\bibitem[\protect\citeauthoryear{{Drinkwater}, {Gregg}, {Hilker}, {Bekki},
  {Couch}, {Ferguson}, {Jones} \& {Phillipps}}{{Drinkwater}
  et~al.}{2003}]{drin03}
{Drinkwater} M.~J.,  {Gregg} M.~D.,  {Hilker} M.,  {Bekki} K.,  {Couch} W.~J.,
  {Ferguson} H.~C.,  {Jones} J.~B.,    {Phillipps} S.,  2003, Nature, 423, 519

\bibitem[\protect\citeauthoryear{{Faifer}, {Forte}, {Norris}, {Bridges},
  {Forbes}, {Zepf}, {Beasley}, {Gebhardt}, {Hanes} \& {Sharples}}{{Faifer}
  et~al.}{2011}]{faif11}
{Faifer} F.~R.,  {Forte} J.~C.,  {Norris} M.~A.,  {Bridges} T.,  {Forbes}
  D.~A.,  {Zepf} S.~E.,  {Beasley} M.,  {Gebhardt} K.,  {Hanes} D.~A.,
  {Sharples} R.~M.,  2011, MNRAS, 416, 155

\bibitem[\protect\citeauthoryear{{Forbes}, {Brodie} \& {Grillmair}}{{Forbes}
  et~al.}{1997}]{forb97}
{Forbes} D.~A.,  {Brodie} J.~P.,    {Grillmair} C.~J.,  1997, AJ, 113, 1652

\bibitem[\protect\citeauthoryear{{Forbes} \& {Forte}}{{Forbes} \&
  {Forte}}{2001}]{forb01}
{Forbes} D.~A.,  {Forte} J.~C.,  2001, MNRAS, 322, 257

\bibitem[\protect\citeauthoryear{{Forbes}, {Ponman} \& {O'Sullivan}}{{Forbes}
  et~al.}{2012}]{forb12}
{Forbes} D.~A.,  {Ponman} T.,    {O'Sullivan} E.,  2012, MNRAS, 425, 66

\bibitem[\protect\citeauthoryear{{Forte}, {Faifer} \& {Geisler}}{{Forte}
  et~al.}{2005}]{fort05}
{Forte} J.~C.,  {Faifer} F.~R.,    {Geisler} D.,  2005, MNRAS, 357, 56

\bibitem[\protect\citeauthoryear{{Forte}, {Faifer} \& {Geisler}}{{Forte}
  et~al.}{2007}]{fort07}
{Forte} J.~C.,  {Faifer} F.~R.,    {Geisler} D.,  2007, MNRAS, 382, 1947

\bibitem[\protect\citeauthoryear{{Forte}, {Faifer}, {Vega}, {Bassino}, {Smith
  Castelli}, {Cellone} \& {Geisler}}{{Forte} et~al.}{2013}]{fort13}
{Forte} J.~C.,  {Faifer} F.~R.,  {Vega} E.~I.,  {Bassino} L.~P.,  {Smith
  Castelli} A.~V.,  {Cellone} S.~A.,    {Geisler} D.,  2013, MNRAS, 431, 1405

\bibitem[\protect\citeauthoryear{{Forte}, {Vega}, {Faifer}, {Smith Castelli},
  {Escudero}, {Gonz\'alez} \& {Sesto}}{{Forte} et~al.}{2014}]{fort14}
{Forte} J.~C.,  {Vega} E.~I.,  {Faifer} F.~R.,  {Smith Castelli} A.~V.,
  {Escudero} C.,  {Gonz\'alez} N.~M.,    {Sesto} L.,  2014, MNRAS, 441, 1391

\bibitem[\protect\citeauthoryear{{Forte}, {Vega} \& {Faifer}}{{Forte}
  et~al.}{2009}]{fort09}
{Forte} J.~C.,  {Vega} I.,    {Faifer} F.~R.,  2009, MNRAS, 397, 1003

\bibitem[\protect\citeauthoryear{{Fukugita}, {Ichikawa}, {Gunn}, {Shimasaku} \&
  {Schneider}}{{Fukugita} et~al.}{1996}]{fuku96}
{Fukugita} M.,  {Ichikawa} T.,  {Gunn} J.,  {Shimasaku} M.,    {Schneider} D.,
  1996, AJ, 111, 1748

\bibitem[\protect\citeauthoryear{{Gebhardt} \& {Kissler-Patig}}{{Gebhardt} \&
  {Kissler-Patig}}{1999}]{gebh99}
{Gebhardt} K.,  {Kissler-Patig} M.,  1999, AJ, 118, 1526

\bibitem[\protect\citeauthoryear{{Georgiev}, {Puzia}, {Goudfrooij} \&
  {Hilker}}{{Georgiev} et~al.}{2010}]{geor10}
{Georgiev} I.~Y.,  {Puzia} T.~H.,  {Goudfrooij} P.,    {Hilker} M.,  2010,
  mnras, 406, 1967

\bibitem[\protect\citeauthoryear{{Gordon}, {Koribalski} \& {Jones}}{{Gordon}
  et~al.}{2003}]{gord03}
{Gordon} S.,  {Koribalski} B.,    {Jones} K.,  2003, MNRAS, 342, 939

\bibitem[\protect\citeauthoryear{{Harris}}{{Harris}}{2001}]{harr01}
{Harris} W.~E.,  2001, in {Labhardt} L.,  {Binggeli} B.,  eds, Saas-Fee
  Advanced Course 28: Star Clusters {Globular cluster systems}.
p.~223

\bibitem[\protect\citeauthoryear{{Harris}}{{Harris}}{2009a}]{harr09}
{Harris} W.~E.,  2009a, ApJ, 699, 254

\bibitem[\protect\citeauthoryear{{Harris}}{{Harris}}{2009b}]{harr09b}
{Harris} W.~E.,  2009b, ApJ, 703, 939

\bibitem[\protect\citeauthoryear{{Harris}, {Harris} \& {Alessi}}{{Harris}
  et~al.}{2013}]{harr13}
{Harris} W.~E.,  {Harris} G.~L.~H.,    {Alessi} M.,  2013, apj, 772, 82

\bibitem[\protect\citeauthoryear{{Harris} \& {van den Bergh}}{{Harris} \& {van
  den Bergh}}{1981}]{harr81}
{Harris} W.~E.,  {van den Bergh} S.,  1981, AJ, 86, 1627

\bibitem[\protect\citeauthoryear{{Ho}, {Li}, {Barth}, {Seigar} \& {Peng}}{{Ho}
  et~al.}{2011}]{ho11}
{Ho} L.~C.,  {Li} Z.-Y.,  {Barth} A.~J.,  {Seigar} M.~S.,    {Peng} C.~Y.,
  2011, ApJS, 197, 19

\bibitem[\protect\citeauthoryear{{Hook}, {J\o rgensen}, {Allington-Smith},
  {Davies}, {Metcalfe}, {Murowinski} \& {Crampton}}{{Hook}
  et~al.}{2004}]{hook04}
{Hook} I.~M.,  {J\o rgensen} I.,  {Allington-Smith} J.~R.,  {Davies} R.~L.,
  {Metcalfe} N.,  {Murowinski} R.~G.,    {Crampton} D.,  2004, PASP, 116, 425

\bibitem[\protect\citeauthoryear{{Hudson}, {Harris} \& {Harris}}{{Hudson}
  et~al.}{2014}]{huds14}
{Hudson} M.~J.,  {Harris} G.~L.,    {Harris} W.~E.,  2014, apjl, 787, L5

\bibitem[\protect\citeauthoryear{{Jacoby}, {Branch}, {Ciardullo}, {Davies},
  {Harris}, {Pierce}, {Pritchet}, {Tonry} \& {Welch}}{{Jacoby}
  et~al.}{1992}]{jaco92}
{Jacoby} G.~H.,  {Branch} D.,  {Ciardullo} R.,  {Davies} R.~L.,  {Harris}
  W.~E.,  {Pierce} M.~J.,  {Pritchet} C.~J.,  {Tonry} J.~L.,    {Welch} D.~L.,
  1992, PASP, 104, 599

\bibitem[\protect\citeauthoryear{{Jedrzejewski}}{{Jedrzejewski}}{1987}]{jedr87}
{Jedrzejewski} R.~I.,  1987, MNRAS, 226, 747

\bibitem[\protect\citeauthoryear{{J\o rgensen}}{{J\o rgensen}}{2009}]{jorg09}
{J\o rgensen} I.,  2009, PASA, 26, 17

\bibitem[\protect\citeauthoryear{{Jord\'an}, {McLaughlin}, {C\^ot\'e},
  {Ferrarese}, {Peng}, {Mei}, {Villegas}, {Merritt}, {Tonry} \&
  {West}}{{Jord\'an} et~al.}{2007}]{jord07}
{Jord\'an} A.,  {McLaughlin} D.~E.,  {C\^ot\'e} P.,  {Ferrarese} L.,  {Peng}
  E.~W.,  {Mei} S.,  {Villegas} D.,  {Merritt} D.,  {Tonry} J.~L.,    {West}
  M.~J.,  2007, ApJ, 171, 101

\bibitem[\protect\citeauthoryear{{Kormendy} \& {Bender}}{{Kormendy} \&
  {Bender}}{1996}]{korm96}
{Kormendy} J.,  {Bender} R.,  1996, ApJ, 464, 119

\bibitem[\protect\citeauthoryear{{Kundu} \& {Whitmore}}{{Kundu} \&
  {Whitmore}}{2001}]{kund01}
{Kundu} A.,  {Whitmore} B.~C.,  2001, AJ, 121, 2950

\bibitem[\protect\citeauthoryear{{Larsen}, {Brodie}, {Huchra}, {Forbes} \&
  {Grillmair}}{{Larsen} et~al.}{2001}]{lars01}
{Larsen} S.~S.,  {Brodie} J.~P.,  {Huchra} J.~P.,  {Forbes} D.,    {Grillmair}
  C.~J.,  2001, AJ, 121, 2974

\bibitem[\protect\citeauthoryear{{Li}, {Ho}, {Barth} \& {Peng}}{{Li}
  et~al.}{2011}]{li11}
{Li} Z.,  {Ho} L.~C.,  {Barth} A.~J.,    {Peng} C.~Y.,  2011, ApJ, 197, 22

\bibitem[\protect\citeauthoryear{{Machacek}, {O'Sullivan}, {Randall}, {Jones}
  \& {Forman}}{{Machacek} et~al.}{2010}]{mach10}
{Machacek} M.~E.,  {O'Sullivan} E.,  {Randall} S.~W.,  {Jones} C.,    {Forman}
  W.~R.,  2010, ApJ, 711, 1316

\bibitem[\protect\citeauthoryear{{Mar\'in-Franch} \&
  {Aparicio}}{{Mar\'in-Franch} \& {Aparicio}}{2002}]{mari02}
{Mar\'in-Franch} A.,  {Aparicio} A.,  2002, ApJ, 568, 174

\bibitem[\protect\citeauthoryear{{McLaughlin}}{{McLaughlin}}{1999}]{mcla99}
{McLaughlin} D.~E.,  1999, aj, 117, 2398

\bibitem[\protect\citeauthoryear{{McLaughlin}, {Harris} \&
  {Hanes}}{{McLaughlin} et~al.}{1994}]{mcla94}
{McLaughlin} D.~E.,  {Harris} W.~E.,    {Hanes} D.~A.,  1994, ApJ, 422, 486

\bibitem[\protect\citeauthoryear{{Mendel}, {Proctor} \& {Forbes}}{{Mendel}
  et~al.}{2007}]{mend07}
{Mendel} J.~T.,  {Proctor} R.~N.,    {Forbes} D.~A.,  2007, MNRAS, 379, 1618

\bibitem[\protect\citeauthoryear{{Mieske}, {Hilker}, {Infante} \&
  {Jord\'an}}{{Mieske} et~al.}{2006}]{mies06}
{Mieske} S.,  {Hilker} M.,  {Infante} L.,    {Jord\'an} A.,  2006, AJ, 131,
  2442

\bibitem[\protect\citeauthoryear{{Mieske}, {Jord\'an}, {C\^ot\'e},
  {Kissler-Patig}, {Peng}, {Ferrarese}, {Blakeslee}, {Mei}, {Merritt}, {Tonry}
  \& {West}}{{Mieske} et~al.}{2006}]{mies06b}
{Mieske} S.,  {Jord\'an} A.,  {C\^ot\'e} P.,  {Kissler-Patig} M.,  {Peng} E.,
  {Ferrarese} L.,  {Blakeslee} J.,  {Mei} S.,  {Merritt} D.,  {Tonry} J.,
  {West} M.,  2006, ApJ, 653, 193

\bibitem[\protect\citeauthoryear{{Mieske}, {Jord{\'a}n}, {C{\^o}t{\'e}},
  {Peng}, {Ferrarese}, {Blakeslee}, {Mei}, {Baumgardt}, {Tonry}, {Infante} \&
  {West}}{{Mieske} et~al.}{2010}]{mies10}
{Mieske} S.,  {Jord{\'a}n} A.,  {C{\^o}t{\'e}} P.,  {Peng} E.~W.,  {Ferrarese}
  L.,  {Blakeslee} J.~P.,  {Mei} S.,  {Baumgardt} H.,  {Tonry} J.~L.,
  {Infante} L.,    {West} M.~J.,  2010, apj, 710, 1672

\bibitem[\protect\citeauthoryear{{Moore}, {Katz}, {Lake}, {Dressler} \&
  {Oemler}}{{Moore} et~al.}{1996}]{moor96}
{Moore} B.,  {Katz} N.,  {Lake} G.,  {Dressler} A.,    {Oemler} A.,  1996,
  Nature, 379, 613

\bibitem[\protect\citeauthoryear{{Muratov} \& {Gnedin}}{{Muratov} \&
  {Gnedin}}{2010}]{mura10}
{Muratov} A.,  {Gnedin} O.,  2010, ApJ, 718, 1266

\bibitem[\protect\citeauthoryear{{O'Sullivan}, {Ponman} \&
  {Collins}}{{O'Sullivan} et~al.}{2003}]{osul03}
{O'Sullivan} E.,  {Ponman} T.~J.,    {Collins} R.~S.,  2003, MNRAS, 340, 1375

\bibitem[\protect\citeauthoryear{{Peng}, {Jord\'an}, {C\^ot\'e}, {Takamiya},
  {West}, {Blakeslee}, {Chin-Wei}, {Ferrarese}, {Mei}, {Tonry} \&
  {West}}{{Peng} et~al.}{2008}]{peng08}
{Peng} E.~W.,  {Jord\'an} A.,  {C\^ot\'e} P.,  {Takamiya} M.,  {West} M.~J.,
  {Blakeslee} J.~P.,  {Chin-Wei} C.,  {Ferrarese} L.,  {Mei} S.,  {Tonry} J.,
   {West} A.,  2008, ApJ, 681, 197

\bibitem[\protect\citeauthoryear{{Perrett}, {Hanes}, {Butterworth},
  {Kavelaars}, {Geisler} \& {Harris}}{{Perrett} et~al.}{1997}]{perr97}
{Perrett} K.~M.,  {Hanes} D.~A.,  {Butterworth} S.~T.,  {Kavelaars} J.,
  {Geisler} D.,    {Harris} W.~E.,  1997, AJ, 113, 895

\bibitem[\protect\citeauthoryear{{Pipino}, {Puzia} \& {Matteucci}}{{Pipino}
  et~al.}{2007}]{pipi07}
{Pipino} A.,  {Puzia} T.~H.,    {Matteucci} F.,  2007, ApJ, 665, 295

\bibitem[\protect\citeauthoryear{{Rejkuba}}{{Rejkuba}}{2012}]{rejk12}
{Rejkuba} M.,  2012, Ap\&SS, 341, 195

\bibitem[\protect\citeauthoryear{{Rusli}, {Thomas}, {Saglia}, {Fabricius},
  {Erwin}, {Bender}, {Nowak}, {Lee}, {Riffeser} \& {Sharp}}{{Rusli}
  et~al.}{2013}]{rusl13}
{Rusli} S.~P.,  {Thomas} J.,  {Saglia} R.~P.,  {Fabricius} M.,  {Erwin} P.,
  {Bender} R.,  {Nowak} N.,  {Lee} C.~H.,  {Riffeser} A.,    {Sharp} R.,  2013,
  AJ, 146, 45

\bibitem[\protect\citeauthoryear{{Salaris} \& {Weiss}}{{Salaris} \&
  {Weiss}}{2002}]{sala02}
{Salaris} M.,  {Weiss} A.,  2002, A\&A, 388, 492

\bibitem[\protect\citeauthoryear{{Schlafly} \& {Finkbeiner}}{{Schlafly} \&
  {Finkbeiner}}{2011}]{schl11}
{Schlafly} E.~F.,  {Finkbeiner} D.~P.,  2011, ApJ, 737, 103

\bibitem[\protect\citeauthoryear{{Schuberth}, {Richtler}, {Hilker}, {Dirsch},
  {Bassino}, {Romanowsky} \& {Infante}}{{Schuberth} et~al.}{2010}]{schu10}
{Schuberth} Y.,  {Richtler} T.,  {Hilker} M.,  {Dirsch} B.,  {Bassino} L.~P.,
  {Romanowsky} A.~J.,    {Infante} L.,  2010, A\&A, 513, 30

\bibitem[\protect\citeauthoryear{{Spitzer} \& {Baade}}{{Spitzer} \&
  {Baade}}{1951}]{spit51}
{Spitzer} L.~J.,  {Baade} W.,  1951, ApJ, 113, 413

\bibitem[\protect\citeauthoryear{{Stetson}}{{Stetson}}{1987}]{stet87}
{Stetson} P.~B.,  1987, PASP, 99, 191

\bibitem[\protect\citeauthoryear{{Strader}, {Brodie}, {Spitler} \&
  {Beasley}}{{Strader} et~al.}{2006}]{stra06}
{Strader} J.,  {Brodie} J.~P.,  {Spitler} L.,    {Beasley} M.,  2006, AJ, 132,
  2333

\bibitem[\protect\citeauthoryear{{Strader} \& {Smith}}{{Strader} \&
  {Smith}}{2008}]{stra08}
{Strader} J.,  {Smith} G.~H.,  2008, AJ, 136, 1828

\bibitem[\protect\citeauthoryear{{Sun}, {Jones}, {Forman}, {Nulsen}, {Donahue}
  \& {Voit}}{{Sun} et~al.}{2006}]{sun06}
{Sun} M.,  {Jones} C.,  {Forman} W.,  {Nulsen} P.~E.,  {Donahue} M.,    {Voit}
  G.~M.,  2006, ApJ, 637, 81

\bibitem[\protect\citeauthoryear{{Tal}, {van Dokkum}, {Nelan} \&
  {Bezanson}}{{Tal} et~al.}{2009}]{tal09}
{Tal} T.,  {van Dokkum} P.~G.,  {Nelan} J.,    {Bezanson} R.,  2009, AJ, 138,
  1417

\bibitem[\protect\citeauthoryear{{Tonini}}{{Tonini}}{2013}]{toni13}
{Tonini} C.,  2013, ApJ, 762, 39

\bibitem[\protect\citeauthoryear{{Tonry}, {Dressler}, {Blakeslee}, {Ajhar},
  {Fletcher}, {Luppino}, {Metzger} \& {Moore}}{{Tonry} et~al.}{2001}]{tonr01}
{Tonry} J.~L.,  {Dressler} A.,  {Blakeslee} J.~P.,  {Ajhar} E.~A.,  {Fletcher}
  A.~B.,  {Luppino} G.~A.,  {Metzger} M.~R.,    {Moore} C.~B.,  2001, ApJ, 546,
  681

\bibitem[\protect\citeauthoryear{{Usher}, {Forbes}, {Brodie}, {Foster},
  {Spitler}, {Arnold}, {Romanowsky}, {Strader} \& {Pota}}{{Usher}
  et~al.}{2012}]{ushe12}
{Usher} C.,  {Forbes} D.~A.,  {Brodie} J.~P.,  {Foster} C.,  {Spitler} L.~R.,
  {Arnold} J.~A.,  {Romanowsky} A.~J.,  {Strader} J.,    {Pota} V.,  2012,
  MNRAS, 426, 1475

\bibitem[\protect\citeauthoryear{{Usher}, {Forbes}, {Spitler}, {Brodie},
  {Romanowsky}, {Strader} \& {Woodley}}{{Usher} et~al.}{2013}]{ushe13}
{Usher} C.,  {Forbes} D.~A.,  {Spitler} L.~R.,  {Brodie} J.~P.,  {Romanowsky}
  A.~J.,  {Strader} J.,    {Woodley} K.~A.,  2013, MNRAS, 436, 1172

\bibitem[\protect\citeauthoryear{{Villegas}, {Jord\'an}, {Peng}, {Blakeslee},
  {John}, {C\^ot\'e}, {Ferrarese}, {Kissler-Patig}, {Mei}, {Infante}, {Tonry}
  \& {West}}{{Villegas} et~al.}{2010}]{vill10}
{Villegas} D.,  {Jord\'an} A.,  {Peng} E.~W.,  {Blakeslee} J.~P.,  {John} P.,
  {C\^ot\'e} P.,  {Ferrarese} L.,  {Kissler-Patig} M.,  {Mei} S.,  {Infante}
  L.,  {Tonry} J.~L.,    {West} M.~J.,  2010, ApJ, 717, 603

\bibitem[\protect\citeauthoryear{{Wegner}, {Bernardi}, {Willmer}, {da Costa},
  {Alonso}, {Pellegrini}, {Maia}, {Chaves} \& {Rit\'e}}{{Wegner}
  et~al.}{2003}]{wegn03}
{Wegner} G.,  {Bernardi} M.,  {Willmer} C.~N.~A.,  {da Costa} L.~N.,  {Alonso}
  M.~V.,  {Pellegrini} P.~S.,  {Maia} M.~A.~G.,  {Chaves} O.~L.,    {Rit\'e}
  C.,  2003, AJ, 126, 2268

\bibitem[\protect\citeauthoryear{{Wehner}, {Harris}, {Whitmore}, {Rothberg} \&
  {Woodley}}{{Wehner} et~al.}{2008}]{wehn08}
{Wehner} E.,  {Harris} W.,  {Whitmore} B.,  {Rothberg} B.,    {Woodley} K.,
  2008, ApJ, 681, 1233

\end{thebibliography}

\end{document}